\documentclass[preprint2]{aastex}
\usepackage{longtable}
\usepackage{rotating}
\usepackage{xcolor}
\usepackage{array}
\usepackage{bigints}
       
\begin{document}
 
 \title{\bf{An investigation of poorly studied open cluster NGC 4337 using multi-color photometric and Gaia~DR2 astrometric data.}}
 
 \author{D. Bisht$^1$, W. H. Elsanhoury$^{2,3}$, Qingfeng Zhu$^1$, Devesh P. Sariya$^4$, R. K. S. Yadav$^5$, Geeta Rangwal$^6$,
         Alok Durgapal$^6$, Ing-Guey Jiang$^4$} 
 
 \affil{
 {$^1$Key Laboratory for Researches in Galaxies and Cosmology, University of Science and Technology of China, Chinese Academy of Sciences,
      Hefei, Anhui, 230026, China}\\
      {email: dbisht@ustc.edu.cn}\\
 {$^2$Astronomy Department, National Research Institute of Astronomy and Geophysics (NRIAG), 11421, Helwan, Cairo, Egypt (Affiliation ID: 60030681)}\\ 
 {$^3$Physics Department, Faculty of Science and Arts, Northern Border University, Turaif Branch, Saudi Arabia}\\ 
 {$^4$Department of Physics and Institute of Astronomy, National Tsing-Hua University, Hsin-Chu, Taiwan}\\ 
 {$^5$Aryabhatta Research Institute of Observational Sciences,Manora Peak, Nainital 263 002, India}\\
 {$^6$Center of Advanced Study, Department of Physics, D. S. B. Campus, Kumaun University Nainital 263002, India}\\ 
}
 
\begin{abstract}
We present a comprehensive analysis (photometric and kinematical) of poorly studied open cluster NGC 4337 using
2MASS, WISE, APASS, and Gaia~DR2 database. By determining the membership probabilities of stars, we identified 624
most probable members with membership probability higher than $50\%$ by using proper motion and parallax data taken
from Gaia~DR2. The mean proper motion of the cluster is obtained as $\mu_{x}=-8.83\pm0.01$ and $\mu_{y}=1.49\pm0.006$ mas yr$^{-1}$.
We find the normal interstellar extinction towards the cluster region. The radial distribution of members provides a cluster radius
of 7.75 arcmin (5.63 pc). The estimated age of $1600\pm180$ Myr indicates that NGC 4337 is an old open cluster with a bunch of red
giant stars. The overall mass function slope for main-sequence stars is found as $1.46\pm0.18$ within the mass range
0.75$-$2.0 $M_\odot$, which is in fair agreement with Salpeter's value (x=1.35) within uncertainty. The present study demonstrates
that NGC 4337 is a dynamically relaxed open cluster. Using the Galactic potential model, Galactic orbits are obtained for
NGC 4337. We found that this object follows a circular path around the Galactic center. Under the kinematical analysis, we compute
the apex coordinates $(A, D)$ by using two methods: (i) the classical convergent point method and (ii) the AD-diagram method. The
obtained coordinates are: $(A_{conv}, D_{conv})$ = (96$^{\textrm{o}}$.27 $\pm$ 0$^{\textrm{o}}$.10,
13$^{\textrm{o}}$.14 $\pm$ 0$^{\textrm{o}}$.27) $\&$ $(A_\circ, D_\circ)$ = (100$^{\textrm{o}}$.282 $\pm$ 0$^{\textrm{o}}$.10,
9$^{\textrm{o}}$.577 $\pm$ 0$^{\textrm{o}}$.323) respectively. We also computed the Velocity Ellipsoid Parameters (VEPs),
matrix elements ($\mu_{ij}$), direction cosines ($l_j$, $m_j$, $n_j$) and the Galactic longitude of the vertex ($l_2$).
\end{abstract}
 
 
\section{Introduction}
\label{Intro}

\begin{figure*}
\begin{center}
\includegraphics[width=8.5cm,height=8.5cm]{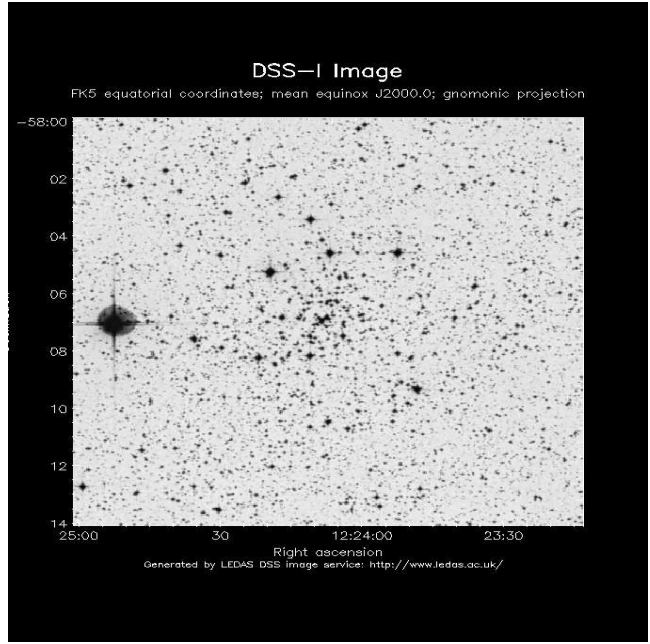}
\caption{The finding chart of stars in NGC 4337, taken from LEDAS.}
\label{id}
\end{center}
\end{figure*}

To understand the stellar evolution and dynamics of stellar systems, open clusters (OCs) are considered as the ideal tools. Before the Gaia era,
the studies of open star clusters were mostly hampered by the presence of field stars. Proper motions and parallaxes information provided
by Gaia's mission in the clusters are very valuable tool to identify the probable members of the clusters. These identified cluster
members are very crucial in deriving the fundamental parameters, mass function and kinematical parameters of the clusters. 

NGC 4337($\alpha_{2000} = 12^{h}24^{m}04^{s}$, $\delta_{2000}=-58^{\circ} 7^{\prime} 24^{\prime\prime}$;
$l$=299$^\circ$.31, $b$=4$^\circ$.55) was previously investigated by Carraro et al. (2014) and Seleznev et al. (2017)
using $UBVI$ photometric and spectroscopic data.
Apart from these studies, no other detailed analysis is available in the literature for this object. Because most OCs are affected
by the field star contamination, the information of cluster membership is necessary to understand cluster properties (Kharchenko et
al. 2013; Cantat-Gaudin et al. 2018). Our main aim of the present study is to perform a deep investigation
of this object using photometric and kinematical database. The Gaia DR2 was made public on 25 April 2018 (Gaia Collaboration
et al. 2016a,b). The data include astrometric five-parameter solutions of central coordinates, proper motion in right ascension
and declination and parallaxes $(\alpha, \delta, \mu_{\alpha}cos\delta, \mu_{\delta}, \pi)$ for more than 1.3 billion
sources (Gaia Collaboration et al., 2018b).

OCs are considered the ideal objects for the determination of initial mass function (IMF, Bisht et al. 2017, 2018, 2019).
In the current paper, one of our motives is to provide useful information about IMF using the most probable cluster members,
which will be helpful to  understand the star formation history in the region of NGC 4337.

In the Solar neighborhood, the characteristic feature of the stellar motion is often represented by the peculiar velocities.
The peculiar velocities have an axis of greatest mobility and this characteristic is represented based on ellipsoidal
law of velocity distribution (Ogorodnikov, 1965). If we consider the ellipsoidal law to be valid at all points within the
steady-state of a stellar system, the function $f$ can be expressed in the form:
$f = F(x, y, z; au^2 + v^2 + cw^2 +2fvw + 2gwu + 2huvb)$, where $a, b, c …, h$ are the functions of $x, y$ and $z$.
In the above generalized formula, the length and distributions of the principal axes of the velocity ellipsoid varies from point
to point in the system. The analysis can be simplified by exploiting the empirical fact that one axis of the velocity ellipsoid
is always found to be oriented perpendicular to the Galactic plane, while the other two axes lie in the plane. It is also found
empirically that the longest axis of the ellipsoid (i.e. the direction of maximum velocity dispersion) points approximately towards
the direction of the Galactic center. Therefore, to specify the orientation of the velocity ellipsoid, we need to determine only the
Galactic longitude along which the principal axis lies. This longitude is called the longitude of the vertex ($l_2$). The importance
of the VEPs lies in their connection to the phase density function, which is one of the most important mathematical functions of
stellar astronomy.

The outline of the paper is as follows. The brief descriptions of the data used are described in Section 2. Section 3 deals with the
study of proper motion and determination of membership probability of stars. The structural properties and derivation of fundamental
parameters using the most probable cluster members have been carried out in Section 4. Luminosity and mass function are discussed
in Section 5. The dynamics and kinematics of the cluster are devoted to Section 6. The conclusions are presented in Section 7.

\section{Data Used}
\label{OBS}

We collected photometric and astrometric data from Gaia DR2 along with the broad-band photometric data sets from
APASS, 2MASS, and WISE for NGC 4337. We cross-matched each catalog for the present study. The description of the data
sets used is as follows.

\subsection{The 2MASS data sets}

For photometric analysis, we used 2MASS data within the 10 arcmin radius of the cluster. The two micros all-sky
($2MASS$, Skrutskie et al. 2006) survey uses two highly automated 1.3m telescopes, one at Mt. Hopkins, Arizona (AZ),
USA and the other at CTIO, Chile with a 3-channel camera (256$\times$256 array of HgCdTe detectors in each channel).
$2MASS$ catalog provides $J~ (1.25~ \mu m)$, $H~ (1.65~ \mu m)$ and $K_{s}~ (2.17~ \mu m)$ band photometry for millions
of galaxies and nearly a half-billion stars (Carpenter, 2001). The sensitivity of this catalog is 15.8 mag for $J$ ,
15.1 mag for $H$ and 14.3 mag for $K_{s}$ band at $S/N$=10. The identification map ofNGC 4337 taken from Leicester database and
archive service (LEDAS) is shown in Fig.\ref{id}. The errors in $J$, $H$ and $K_{s}$ bands are
plotted against $J$ magnitude in the upper left panel of Fig.~\ref{error}. This figure shows that the mean error in $J$,
$H$ and $K_{s}$ band is $\le$ 0.04 at $J\sim$ 13 mag. The error becomes $\sim$0.1 at $J$$\sim$16 mag.

\subsection{WISE data}

The effective wavelength of Wide-field Infrared Survey Explorer (WISE; Wright et al. 2010) are $3.35 \mu m$ (W1), $4.60 \mu m $(W2),
$11.56 \mu m$ (W3) and $22.09 \mu m$ (W4) in the mid-IR bands. We used ALLWISE source catalog to extract data for NGC 4337.
This catalog has achieved $5\sigma$ point source sensitivities better than 0.08, 0.11, 1 and 6 mJy at 3.35, 4.60, 11.56
and 22.09 $\mu m$, which is expected to be more than $99\%$ of the sky. These sensitivities are 16.5, 15.5, 11.2 and 7.9 mag
for W1, W2, W3 and W4 bands correspond to vega magnitudes. The photometric errors for $W1$, $W2$
and $W3$ bands are plotted against $J$ magnitude in the upper left panel of Fig.~\ref{error}.

\subsection{Gaia DR2 data sets}

\begin{figure*}
\begin{center}
\hbox{
\includegraphics[width=8.5cm, height=8.5cm]{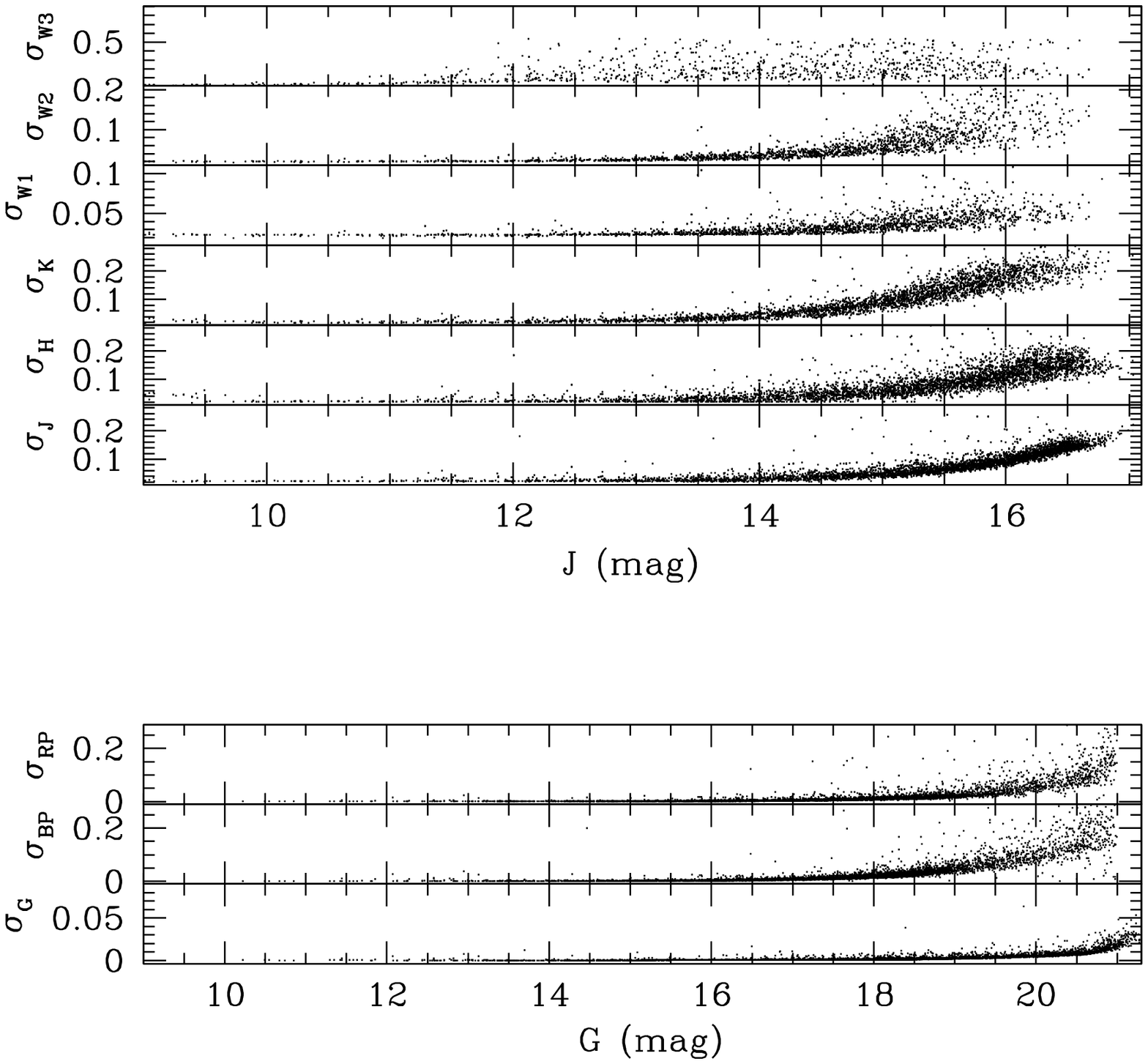}
\includegraphics[width=8.5cm, height=8.5cm]{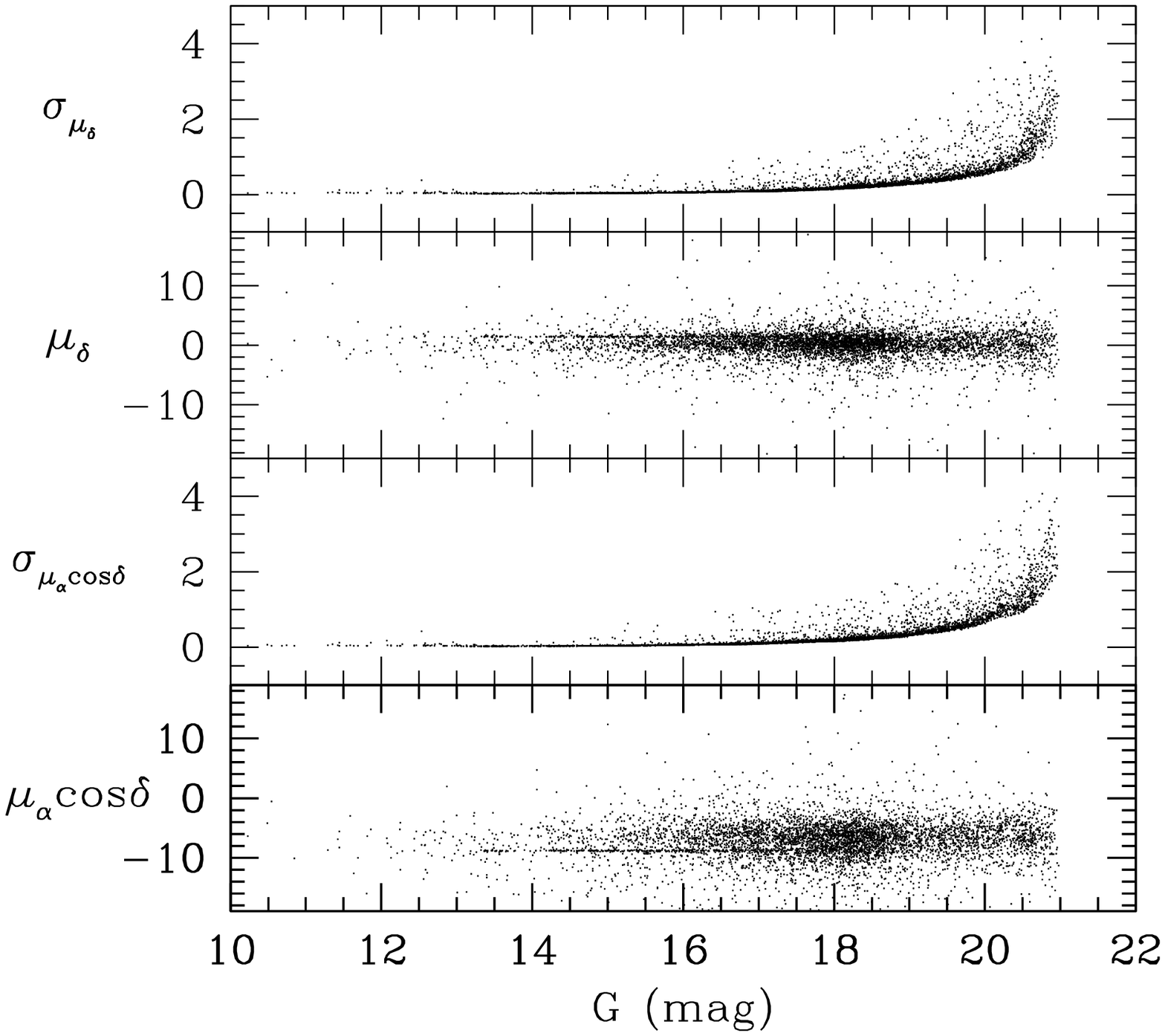}
}
\vspace{-0.5cm}
\caption{(Left upper panels) Photometric errors in $2MASS$ $J$, $H$ and $K_{s}$ band against $J$ magnitude. (Left lower panels)
Photometric errors in Gaia bands ($G$, $G_{BP}$ and $G_{RP}$) with $G$ magnitudes. (Right panels) Plot of proper motion and their
respective errors versus $G$ magnitude.}
\label{error}
\end{center}
\end{figure*}

Gaia~DR2 (Gaia Collaboration et al. 2018a) database within 10 arcmin radius of the cluster are used here for the astrometric
study of NGC 4337. This data consist of positions on the sky $(\alpha, \delta)$, parallaxes and proper motions
($\mu_{\alpha} cos\delta , \mu_{\delta}$) with a  limiting magnitude of $G=21$ mag. The photometric errors in Gaia passbands 
($G$, $G_{BP}$ and $G_{RP}$) versus $G$ magnitudes are plotted in the lower-left panel of Fig.~\ref{error}. The uncertainties
in parallax values are $\sim$ 0.04 milliarcsecond (mas) for sources at $G\le15$ mag and $\sim$ 0.1 mas for sources with $G\sim17$ mag.
The proper motions with their respective errors are plotted against $G$ magnitude in the right panel of Fig.~\ref{error}.
The uncertainties in the corresponding proper motion components are $\sim$ 0.06 mas $yr^{-1}$ (for $G\le15$ mag), $\sim$0.2
mas $yr^{-1}$ (for $G\sim17$ mag) and $\sim$1.2 mas $yr^{-1}$ (for $G\sim20$ mag).

\subsection{\bf APASS data}

The American Association of Variable Star Observers (AAVSO) Photometric All-Sky Survey (APASS) is organized in five
filters: $B$, $V$ (Landolt) and $g^{\prime}$, $r^{\prime}$, $i^{\prime}$ proving stars with $V$ band magnitude range
from 7 to 17 mag (Heden \& Munari 2014). Their latest catalog DR9 covers almost $99\%$ sky (Heden et al. 2016).
For NGC 4337, we extracted APASS data from $http://vizier.u-strasbg.fr/viz-bin/VizieR?-source=II/336$.

\begin{figure*}
\begin{center}
\includegraphics[width=15.5cm, height=15.5cm]{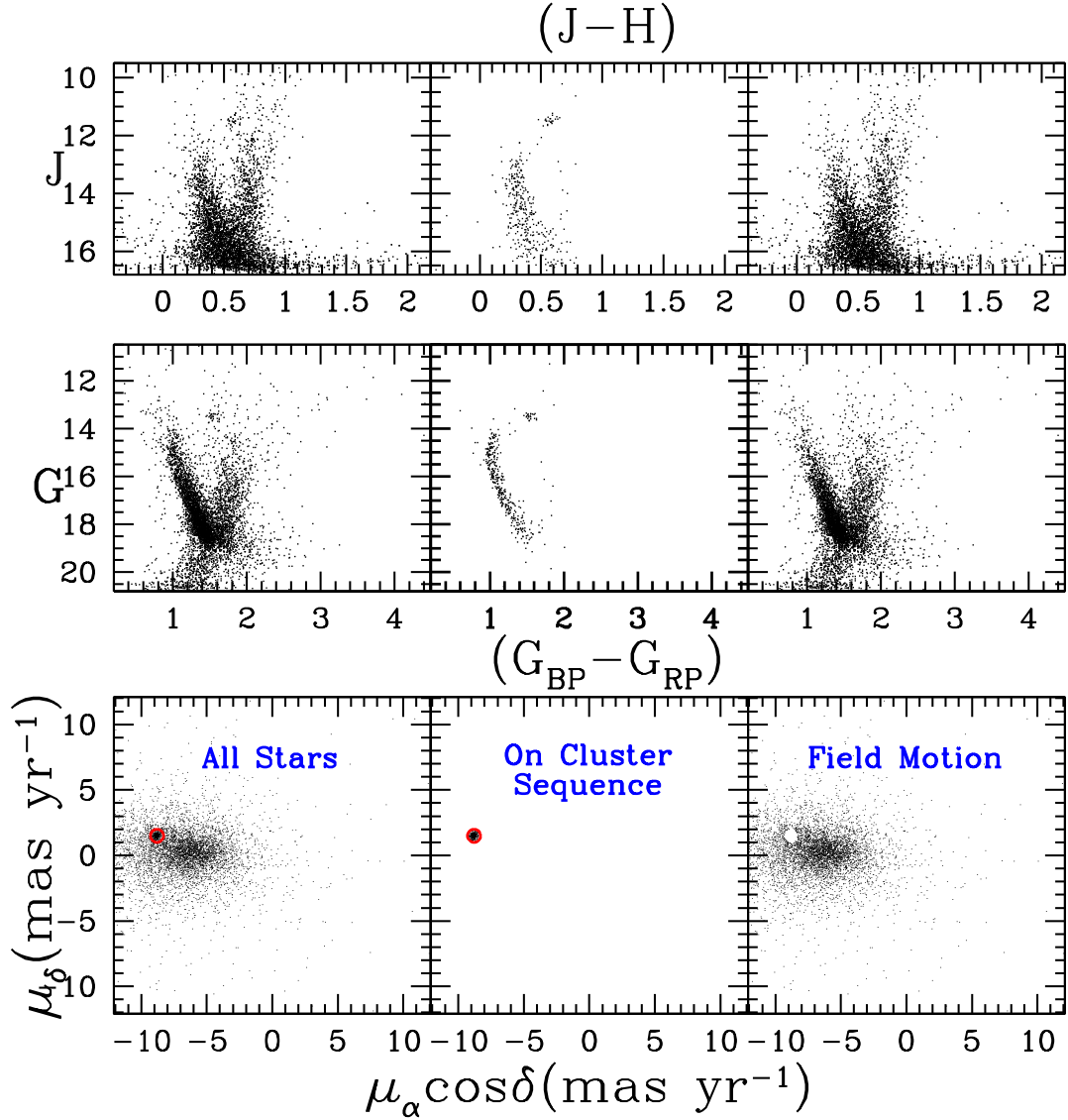}
\caption{( Bottom panels) Proper-motion vector point diagram (VPD) for NGC 4337 based on Gaia~DR2. (Top panels) $J$ versus
$J-H$ color magnitude diagrams. (Middle panels) $G$ versus $(G_{BP}-G_{RP})$ color magnitude diagrams. (Left) The entire sample.
(Center) Stars in VPDs within circle of $0.5~ mas~ yr^{-1}$ of the cluster mean. (Right) Probable background/foreground field stars
in the direction of these clusters. We have used only stars with PM $\sigma$ smaller than $0.5~ mas~ yr^{-1}$ in each coordinate.}
\label{pm_hist}
\end{center}
\end{figure*}

\section{Mean proper motion and Field star separation}

Proper motions (PMs) play a vital role in order to separate field stars from the main sequence and to derive authentic
fundamental parameters as well (Yadav et al. 2013; Sariya et al. 2018a; Bisht et al. 2019). Proper motion and parallax are
very reliable parameters to extract field stars from the cluster zone, because cluster stars share the same kinematical properties
and distances (Rangwal et al. 2019). To detect the distribution of cluster and field stars, PM components ($\mu_{\alpha} cos{\delta}$,
$\mu{\delta}$) are plotted as VPD in the bottom panels of Fig.~\ref{pm_hist}. The panels of top rows show the corresponding
$J$ versus $J-H$ and middle row of panels show $G$ versus $(G_{BP}-G_{RP})$ color-magnitude diagrams (CMDs). The left panel
shows all the stars, while the middle and right panels show the probable cluster members and field region stars. A circle of
0.5 mas $yr^{-1}$ around the distribution of cluster stars in the VPD characterize our membership criteria. The chosen radius
in such VPD is a compromise between losing cluster members with poor PMs and including field region stars
(Sariya et al. 2017, 2018b). We have also used mean parallax for the cluster member selection. We estimated the weighted mean
of parallax for stars inside the circle of VPD having $G$ mag brighter than 20$^{th}$ mag. We considered a star as the most probable members
if it lies within 0.5 mas yr$^{-1}$ radius in VPD and has a parallax within 3$\sigma$ from the mean parallax of the cluster.
The CMD of the most probable cluster members is shown in the upper-middle panel in Fig.~\ref{pm_hist}. In this figure, the main
sequence of the cluster is identified. These stars have a PM error of $\le 0.5 ~$ mas yr$^{-1}$.

To estimate the mean proper motion, we consider probable cluster members selected from VPD and CMD as shown in
Fig.~\ref{pm_hist}. We constructed the histograms for $\mu_{\alpha} cos{\delta}$ and $\mu_{\delta}$ as shown in
the left panel of Fig.~\ref{pm_hist1}. The fitting of a Gaussian function to the histograms provides mean proper motion
in both directions. In this way, we found the mean-proper motion of NGC 4337 as $-8.83\pm0.01$ and $1.49\pm0.006$ mas yr$^{-1}$ in
$\mu_{\alpha} cos{\delta}$ and $\mu_{\delta}$ respectively. The estimated values of mean PMs for this object are in very good
agreement with Cantat-Gaudin et al. (2018).

\subsection{Membership probabilities}
\label{MP}

It is essential to identify the most probable cluster members towards the cluster zone for the reliable determination
of its fundamental parameters. Vasilevskis et al. (1958) built up a mathematical technique to estimate
the membership probabilities using PMs. Sanders (1971b) introduced the maximum likelihood principle for the evaluation of
membership of stars in clusters zone. In this paper, we adopted the approach given by Balaguer-N\'{u}\~{n}ez et al. (1998)
by using Gaia PM database. In this method, two frequency distribution functions are set up for a particular $i^{th}$ star.
The frequency distributions of cluster members ($\phi_c^{\nu}$) and field stars ($\phi_f^{\nu}$) are presented by the equations
given below:\\

\begin{figure*}
\begin{center}
\hbox{
\includegraphics[width=8.5cm, height=8.5cm]{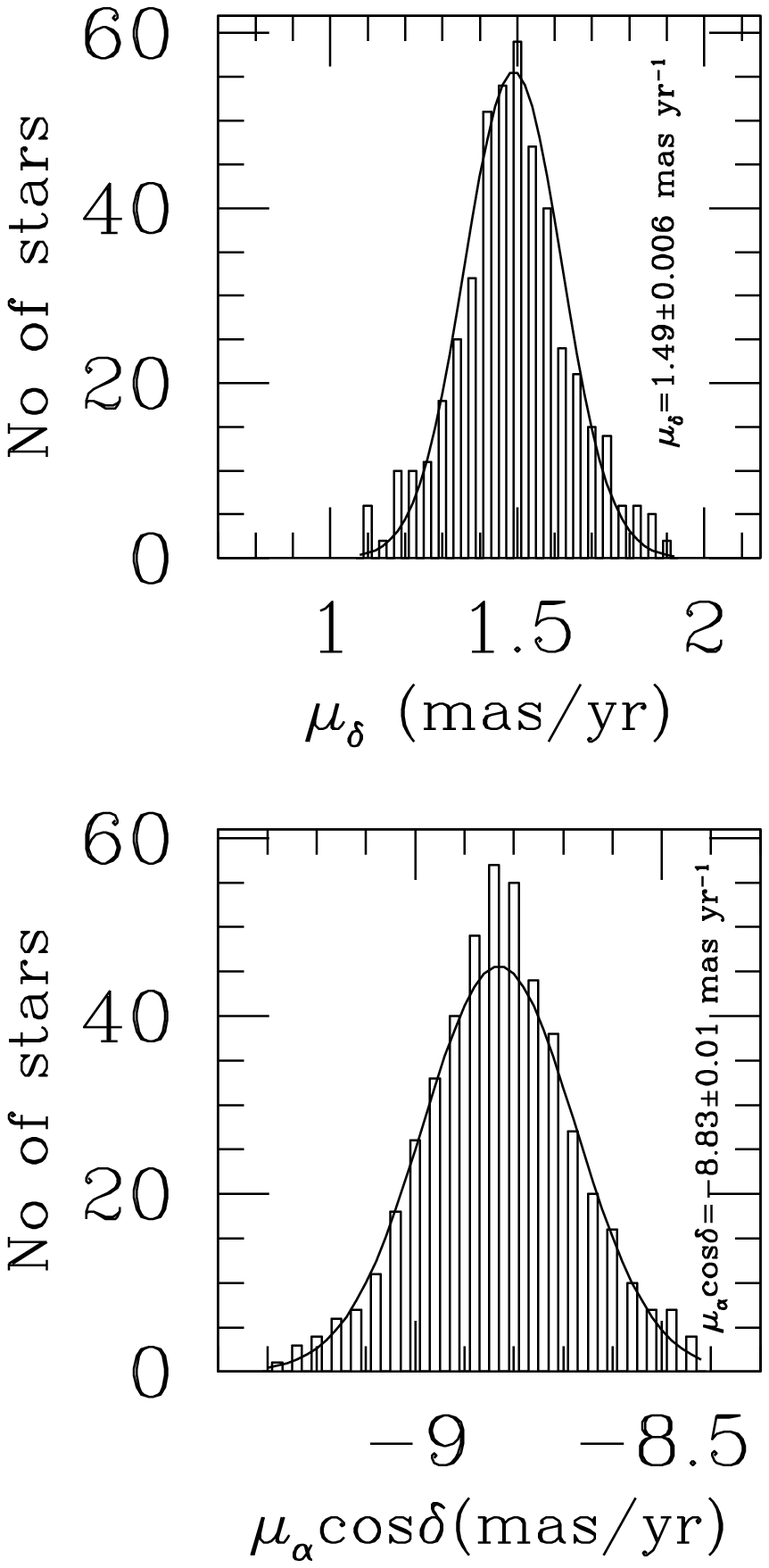}
\hspace{-2cm}\includegraphics[width=8.5cm, height=8.5cm]{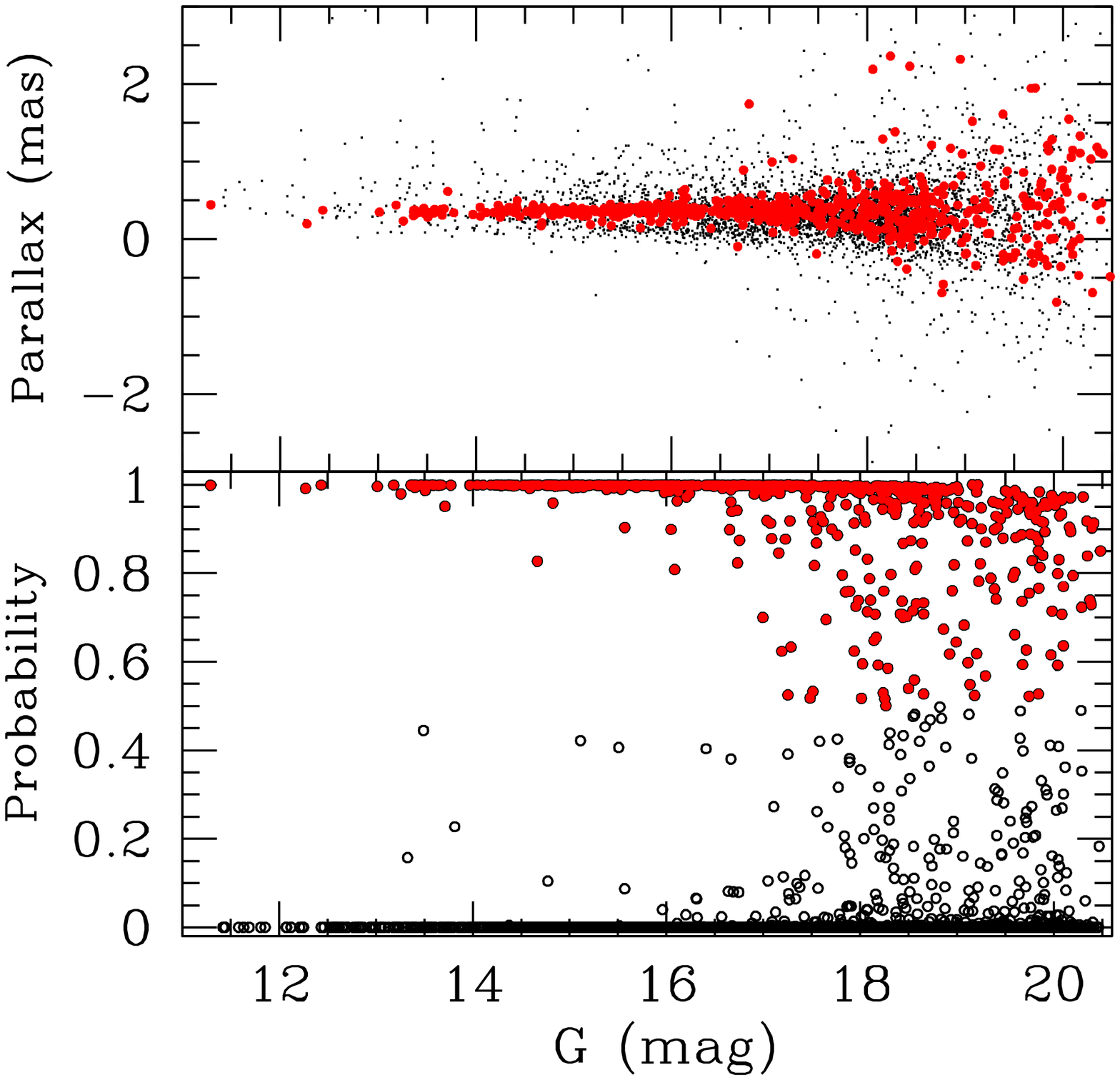}
}
\vspace{-0.5cm}\caption{(Left panels) Histogram for the determination of mean values of proper motion in RA and DEC directions.
The Gaussian function fits to the central bins provide the mean values in both directions as shown in each panel.
(Right panel) Membership probability and parallax values as a function of $G$ magnitude for stars in cluster area. Red
dots have membership probability greater than 50 $\%$.}
\label{pm_hist1}
\end{center}
\end{figure*}
\begin{figure*}
\begin{center}
\hbox{
\includegraphics[width=5.5cm, height=5.5cm]{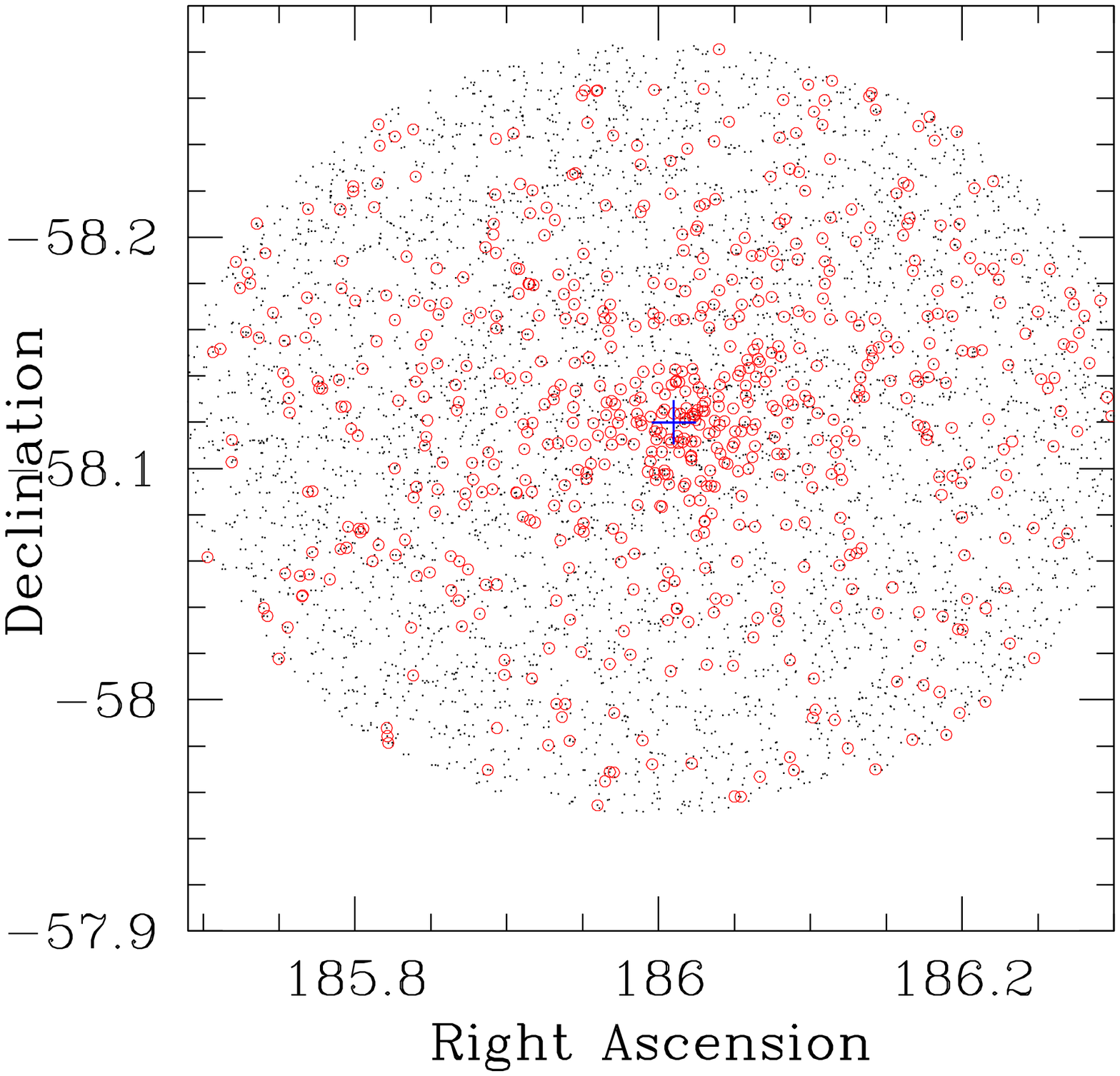}
\includegraphics[width=5.5cm, height=5.5cm]{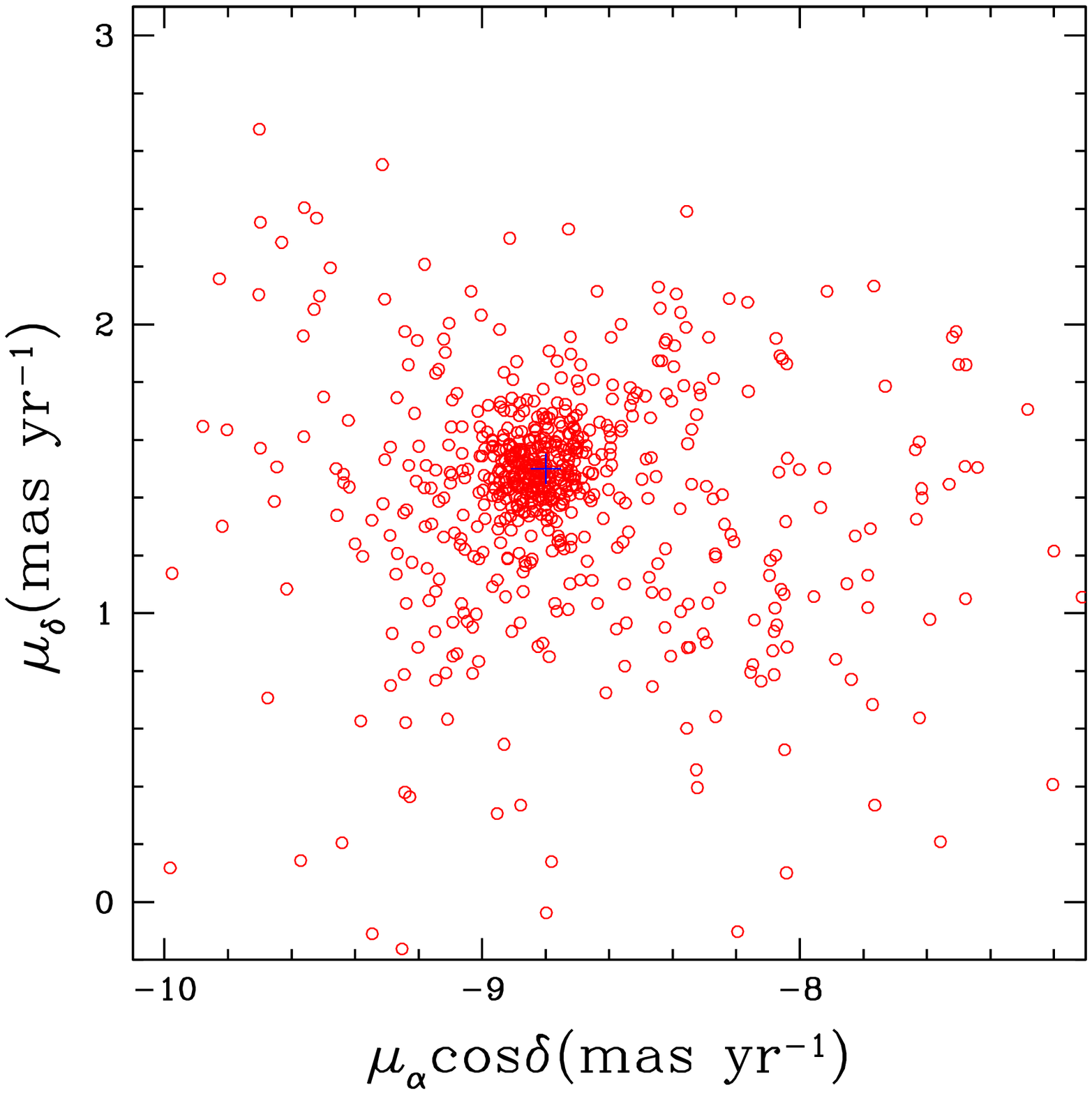}
\includegraphics[width=5.5cm, height=5.5cm]{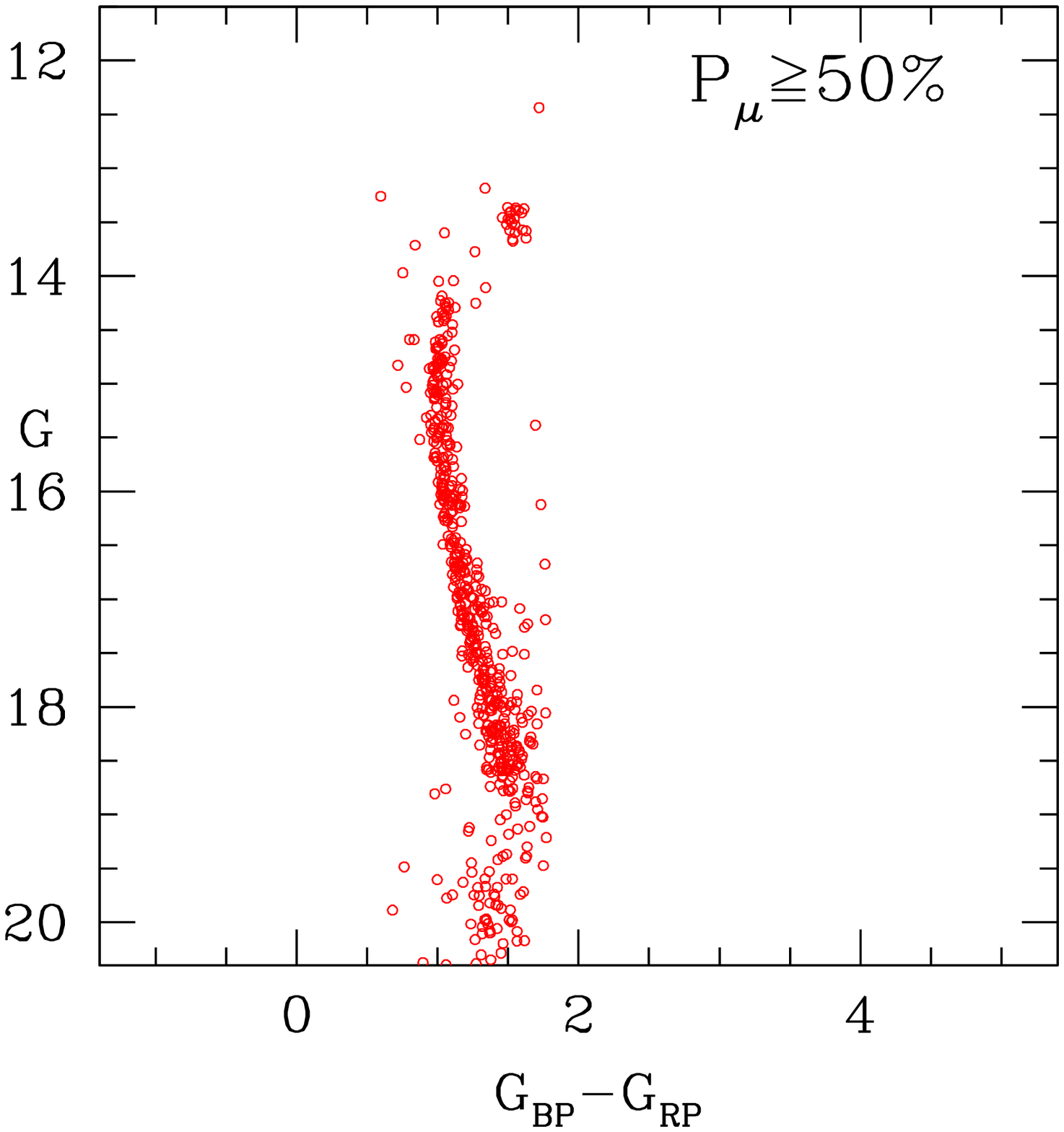}
}
\caption{(Left panel) Identification chart of the cluster. Plus sign indicates cluster center. (Middle panel) Proper motion distribution
of stars towards the cluster region. (Right panel) $G$ versus $G_{BP}-G_{RP}$ color magnitude diagram. Red dots are the most probable
cluster members with membership probability higher than 50 $\%$.}
\label{members}
\end{center}
\end{figure*}
\begin{equation}
\resizebox{0.5\textwidth}{!}{$\phi_c^{\nu} =\frac{1}{2\pi\sqrt{{(\sigma_c^2 + \epsilon_{xi}^2 )} {(\sigma_c^2 + \epsilon_{yi}^2 )}}} \\
    \times exp\{{-\frac{1}{2}[\frac{(\mu_{xi} - \mu_{xc})^2}{\sigma_c^2 + \epsilon_{xi}^2 } + \frac{(\mu_{yi} - \mu_{yc})^2}{\sigma_c^2 + \epsilon_{yi}^2}] }\}$}
\end{equation}

\begin{center}
and\\
\end{center}

\begin{equation}
\resizebox{0.5\textwidth}{!}{$\phi_f^{\nu} =\frac{1}{2\pi\sqrt{(1-\gamma^2)}\sqrt{{(\sigma_{xf}^2 + \epsilon_{xi}^2 )} {(\sigma_{yf}^2 + \epsilon_{yi}^2 )}}}
\\ \times exp\{{-\frac{1}{2(1-\gamma^2)}[\frac{(\mu_{xi} - \mu_{xf})^2}{\sigma_{xf}^2 + \epsilon_{xi}^2}}
-\frac{2\gamma(\mu_{xi} - \mu_{xf})(\mu_{yi} - \mu_{yf})} {\sqrt{(\sigma_{xf}^2 + \epsilon_{xi}^2 ) (\sigma_{yf}^2 + \epsilon_{yi}^2 )}} + \frac{(\mu_{yi} - \mu_{yf})^2}{\sigma_{yf}^2 + \epsilon_{yi}^2}]\}
$}
\end{equation}

where ($\mu_{xi}$, $\mu_{yi}$) are the PMs of $i^{th}$ star, although ($\epsilon_{xi}$, $\epsilon_{yi}$) are corresponding errors
in PMs. ($\mu_{xc}$, $\mu_{yc}$) represent the cluster's PM center and ($\mu_{xf}$, $\mu_{yf}$) are PM center coordinates for
field stars. The intrinsic proper motion dispersion is denoted by $\sigma_c$ for members, whereas $\sigma_{xf}$ and $\sigma_{yf}$
show the field intrinsic proper motion dispersions. The correlation coefficient $\gamma$ is calculated as:\\

\begin{equation}
\gamma = \frac{(\mu_{xi} - \mu_{xf})(\mu_{yi} - \mu_{yf})}{\sigma_{xf}\sigma_{yf}}.
\end{equation}

To figure out $\phi_c^{\nu}$ and $\phi_f^{\nu}$, we considered only those stars which have PM errors better than $\sim$0.5 mas~yr$^{-1}$.
A tight bunch of stars is found at $\mu_{xc}$=$-$8.83 mas~yr$^{-1}$, $\mu_{yc}$=1.49 mas~yr$^{-1}$ and in the circular region having radii of
0.5 mas~yr$^{-1}$. Assuming a distance of 2.5 kpc and the radial velocity dispersion of 1 km $s^{-1}$ for open star clusters
(Girard et al. 1989), the expected dispersion ($\sigma_c$) in PMs would be 0.08 $mas~yr^{-1}$. For field region stars, we have
estimated ($\mu_{xf}$, $\mu_{yf}$) = ($-$6.5, 0.5) mas yr$^{-1}$ and ($\sigma_{xf}$, $\sigma_{yf}$) = (4.5, 3.8) mas yr$^{-1}$.\\

Considering $n_{c}$ and $n_{f}$ are the normalized number of cluster and field stars respectively (i.e., $n_c + n_f = 1$),
the absolute distribution function can be computed as:\\

\begin{equation}
\phi = (n_{c}~\times~\phi_c^{\nu}) + (n_f~\times~\phi_f^{\nu}),  \\
\end{equation}

Finally, the  membership probability of the $i^{th}$ star is given by:\\
\begin{equation}
P_{\mu}(i) = \frac{\phi_{c}(i)}{\phi(i)}. \\
\end{equation}

From this analysis, we identified 624 stars as cluster members with membership probability higher than  $50\%$ and $G\le20$ mag.
In the lower right panel of Fig.~\ref{pm_hist1}, we plotted membership probability versus $G$ magnitude. In this figure, cluster
members and field stars are separated. In the upper right panel of this figure, we plotted $G$ magnitude versus
parallax of stars. In Fig.~\ref{members}, we plotted identification chart in the left panel, proper motion distribution in
the middle panel and $G$ versus $B_{P}-R_{P}$ color-magnitude diagram in the right panel using the most probable cluster members.
The most probable cluster members with high membership probability  $(\ge50\%)$ are shown by red dots in Fig.~\ref{pm_hist1} and
Fig.~\ref{members}.

\begin{figure*}
\begin{center}
\hbox{
\includegraphics[width=7.5cm, height=7.5cm]{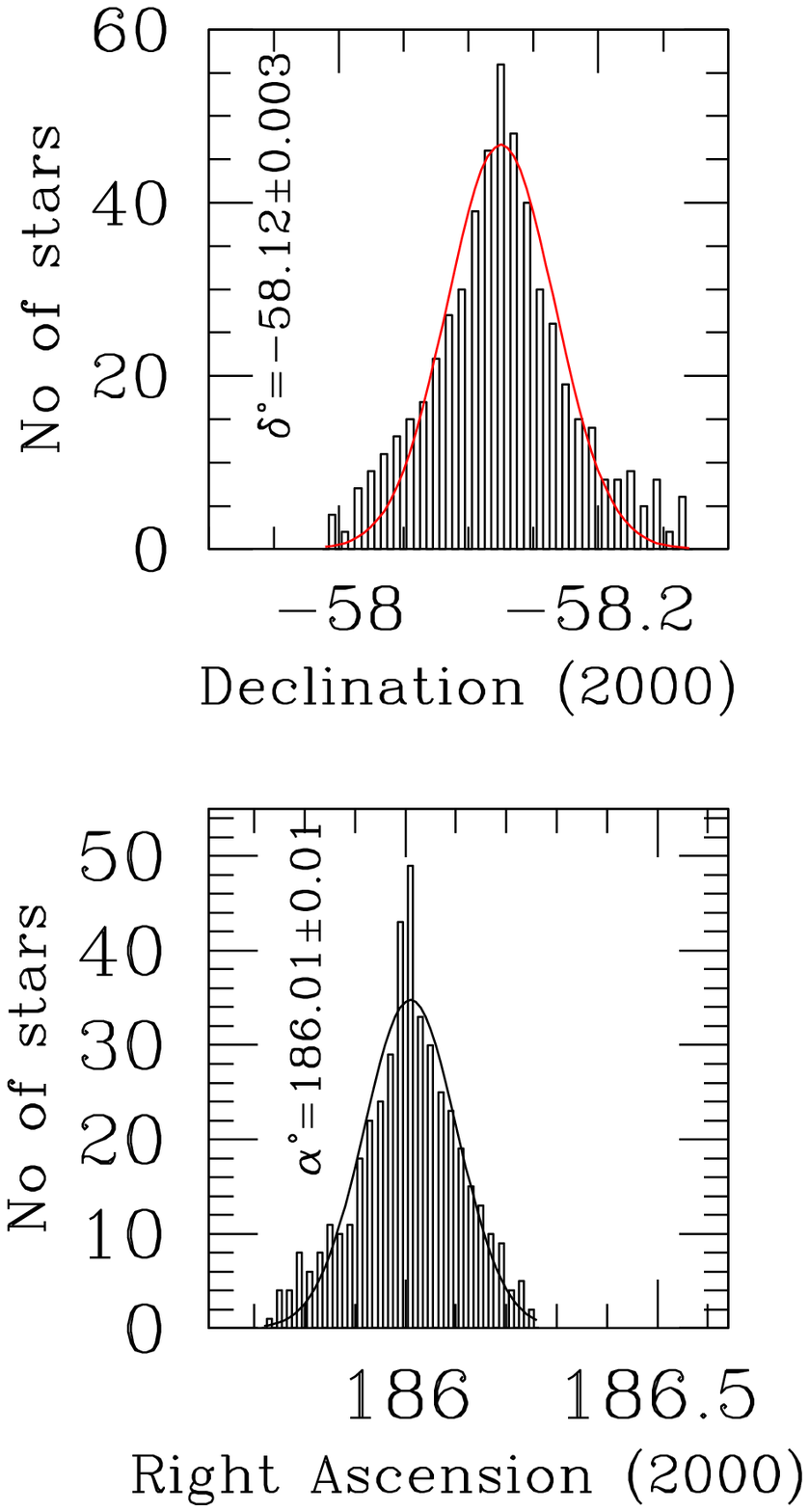}
\hspace{-2.5cm}\includegraphics[width=7.5cm, height=7.5cm]{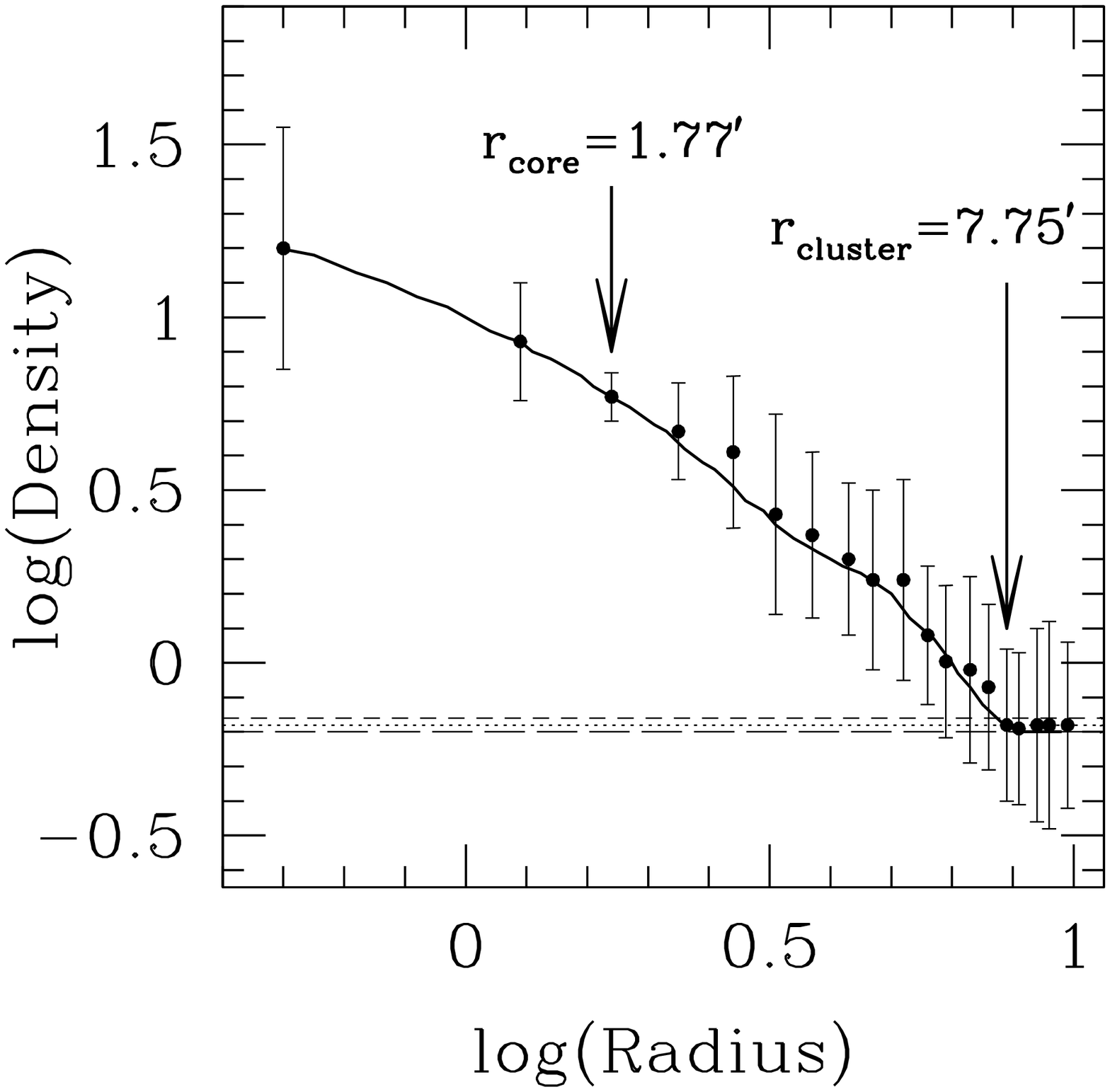}
}
\vspace{-0.5cm}\caption{(Left panel) Profiles of stellar counts across cluster NGC 4337. The Gaussian fits have been applied.
The center of symmetry about the peaks of Right Ascension and Declination is taken to be the position of the cluster
centers. (Right panel) Surface density distribution (log(radius) versus log(density)) of the cluster NGC 4337.
Errors are determined from sampling statistics $\frac{1}{\sqrt{N}}$, where $N$ is the number of stars used in the density
estimation at that point. The smooth line represent the fitted profile whereas dotted line shows the background density level.
Long and short dash lines represent the errors in background density.}
\label{center_dens} 
\end{center}
\end{figure*}

\section{Structural properties of NGC 4337}

\subsection{Spatial structure: radial density profile}

Accurate information of the central coordinates of a cluster is very necessary for a reliable estimation of the cluster's
fundamental parameters, such as age, distance, reddening, etc. For calculating the center, we applied the star-count method
to the whole area of NGC 4337. The resulting histograms in both the RA and DEC directions are shown in the left panel of
Fig.~\ref{center_dens}. The Gaussian curve-fitting is applied at the central regions in the histograms. The fitting provides
the central coordinates as $\alpha = 186.01\pm0.01$ deg ($12^{h} 24^{m} 2.3^{s}$) and $\delta = -58.12\pm0.003$ deg
($-58^{\circ} 7^{\prime} 12^{\prime\prime}$). These values of the cluster center are in agreement with the values given
by Dias et al. (2002). The central coordinates are also very close to the values given by Cantat-Gaudin et al. (2018) for
NGC 4337.

To know about the extent of the cluster, we plotted the radial density profile (RDP) (log(radius) versus log(density))
as shown in the right panel of Fig.~\ref{center_dens} using the derived central coordinates in the above paragraph of this
section. To do this, we divided the area of NGC 4337 into many concentric rings. The number density, $R_{i}$, in
the i$^{th}$ zone is determined by using the formula $R_{i}$ = $\frac{N_{i}}{A_{i}}$, where $N_{i}$ is the number of stars and $A_{i}$ is
the area of the $i^{th}$ zone. This RDP flattens at $r\sim$ 7.75 armin (log(radius)=0.89) and begins to merge with
the background density as shown in the right panel of Fig.~\ref{center_dens}. Therefore, we consider 7.75 arcmin as the
cluster radius. A smooth continuous line represents fitted King (1962) profile:\\

\begin{equation}
f(r) = f_{bg}+\frac{f_{0}}{1+(r/r_{c})^2}\\
\end{equation}

where $r_{c}$, $f_{0}$, and $f_{bg}$ are the core radius, central density, and the background density level, respectively.
By fitting the King model to the radial density profile, we estimated the structural parameters for NGC 4337. The obtained values
of central density, background density, and core radius are: $13.77\pm2.0$ stars per arcmin$^{2}$, $0.41\pm0.11$
stars per arcmin$^{2}$ and $1.77\pm0.2$ arcmin respectively. The background density level with errors is also shown by the dotted
lines. The cluster limiting radius, $r_{lim}$, is calculated  by comparing $f(r)$ to the border background density level, $f_{b}$ ,
defined as:\\

\begin{figure}
\begin{center}
\includegraphics[width=8.5cm, height=8.5cm]{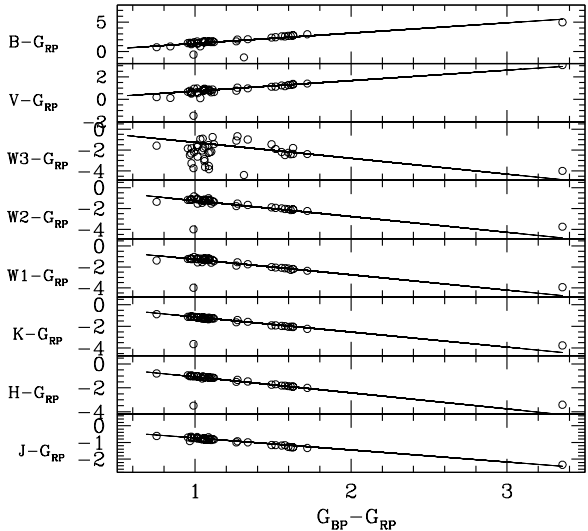}
\vspace{-1cm}\caption{The $(\lambda-G_{RP})/(G_{BP}-G_{RP})$ two color diagrams using the stars selected from VPD. The continuous
lines represent the slope determined through least-squares linear fit.}
\label{cc2}
\end{center}
\end{figure}
\begin{figure}
\begin{center}
\includegraphics[width=8.5cm, height=8.5cm]{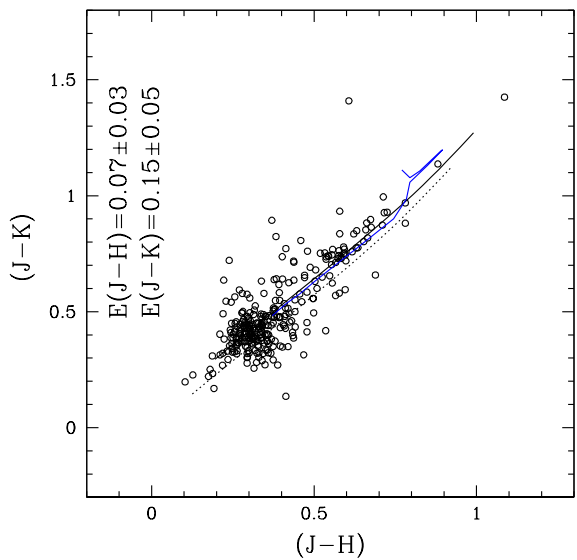}
\vspace{-1cm}\caption{$(J-H)$ vs $(J-K)$ two color diagram. Solid and dotted lines are the ZAMS taken from
Caldwell et al. (1993). Blue line is the theoretical isochrone is taken from Marigo et al. (2017).}
\label{cc_jhjk}
\end{center}
\end{figure}
\begin{equation}
f_{b}=f_{bg}+3\sigma_{bg}\\
\end{equation}

where $\sigma_{bg}$ is uncertainty of $f_{bg}$ . Therefore, $r_{lim}$ is calculated according to the following formula
(Bukowiecki et al. 2011):\\

\begin{equation}
r_{lim}=r_{c}\sqrt(\frac{f_{0}}{3\sigma_{bg}}-1)\\
\end{equation}

The estimated value of limiting radius is found to be 9.66 arcmin. By combining $r_{c}$ and $r_{lim}$ both in terms
of the concentration parameter, $c = log (\frac{r_{lim}} {r_{c}}$, Peterson \& King, 1975), we can characterize the
structure of cluster in the Milky Way. In the present study, the concentration parameter is found to be 0.73.
Maciejewski \& Niedzielski (2007) reported that $R_{lim}$ may vary for individual clusters from 2$R_{c}$ to 7$R_{c}$.
The estimated value of $R_{lim}$ ($\sim$ 5.5$R_{c}$) shows a good agreement with Maciejewski \& Niedzielski (2007).

\subsection{Tidal radius}

Tidal interactions are crucial to understand the initial structure and dynamical evolution of
the clusters (Chumak et al. 2010). Tidal radius is the distance from cluster center where gravitational
acceleration caused by the cluster becomes equal to the tidal acceleration due to parent Galaxy (von Hoerner 1957).
The Galactic mass $M_{G}$ inside a Galactocentric radius $R_{G}$ is given by (Genzel \& Townes, 1987),\\

\begin{equation}
M_{G}=2\times10^{8} M_{\odot} (\frac{R_{G}} {30 pc})^{1.2}\\
\end{equation}

Estimated values of Galactic mass inside the Galactocentric radius (see Sec. 4.5) are found as $1.4\times10^{11} M_{\odot}$.
Kim et al. (2000) has introduced the formula for the tidal radius $R_{t}$ of clusters as \\

\begin{equation}
R_{t}=(\frac{M_{c}} {2M_{G}})^{1/3}\times R_{G}\\
\end{equation}

\begin{table}
\centering
\caption{Multi-band color excess ratios in the direction of NGC 4337.
}
\begin{center}
\small
\begin{tabular}{ccc}
\hline\hline
Band $(\lambda)$ & Effective wavelength  &   $\frac{\lambda-G_{RP}}{G_{BP}-G_{RP}}$ \\
&              (nm)
\\
\hline\hline
Johnson~ B        &445              &$1.70\pm0.03$ \\ 
Johnson~ V        &551              &$0.93\pm0.02$ \\ 
2MASS~ J          &1234.5           &$-0.82\pm0.03$ \\
2MASS~ H          &1639.3           &$-1.14\pm0.03$ \\
2MASS~ K          &2175.7           &$-1.27\pm0.06$ \\
WISE ~W1          &3317.2           &$-1.31\pm0.07$ \\
WISE~ W2          &4550.1           &$-1.50\pm0.08$ \\
WISE~ W3          &12082.3          &$-1.46\pm0.07$ \\
\hline
\end{tabular}
\label{gaia_slope}
\end{center}
\end{table}
where $R_{t}$ and  $M_{c}$ indicate the cluster's tidal radius and total mass (see Sect.~8), respectively. The estimated
value of the tidal radius is $9.78\pm0.62$ pc for NGC 4337.

\subsection{Extinction law towards cluster region}

We combined multi-wavelength photometric data with Gaia DR2  astrometric data to test the extinction law from
optical to mid-infrared region. The resultant $(\lambda-G_{RP})/(G_{BP}-G_{RP})$ two color diagrams
(TCDs) are shown in the right panel of Fig.~\ref{cc2}. Here, $\lambda$ stand for the filters other than
$G_{RP}$. All stars shown in this figure are the most probable cluster members. A linear fit to the data points
is performed and slopes are listed in Table \ref{gaia_slope}. The estimated values of slopes are in good agreement
with the value given by Bisht et al. (2019). We calculated $\frac{A_{V}}{E(B-V)}$ as $\sim$ 3.2. This indicates
that reddening law is normal in the direction of NGC 4337.

\subsection{The reddening using $2MASS$ colors}

Interstellar reddening of the clusters is very important for a reliable determination of the distance and age. To determine
the color-excess, we plotted $(J-H)$ versus $(J-K)$ color-color diagram as shown in Fig.~\ref{cc_jhjk} for NGC 4337.
Stars used in this figure are the most likely members based on VPD. The Zero Age Main Sequence (ZAMS) shown by the
solid line is taken from Caldwell et al. (1993). The same ZAMS showed by the dotted line is shifted by $E(J-H) = 0.07\pm0.03$ mag
and $E(J-K) = 0.15\pm0.05$ mag. The ratio $E(J-H)/E(J-K)$ is found to be 0.46, which is a little bit lower than the normal
interstellar extinction value of 0.55 as suggested by Cardelli et al. (1989). We have determined the value of interstellar
reddening $E(B-V)$ using the relationship $\frac{E(J-K)}{E(B-V)}=0.72\pm0.05$, as given by Morgan \& Nandy (1982). The value
of $E(B-V)$ is found as $0.21\pm0.01$. The present value of interstellar reddening is in very good agreement with
Carraro et al. (2014).

\subsection{Age and distance}

\begin{figure*}
\begin{center}
\hbox{
\includegraphics[width=8.5cm, height=8.5cm]{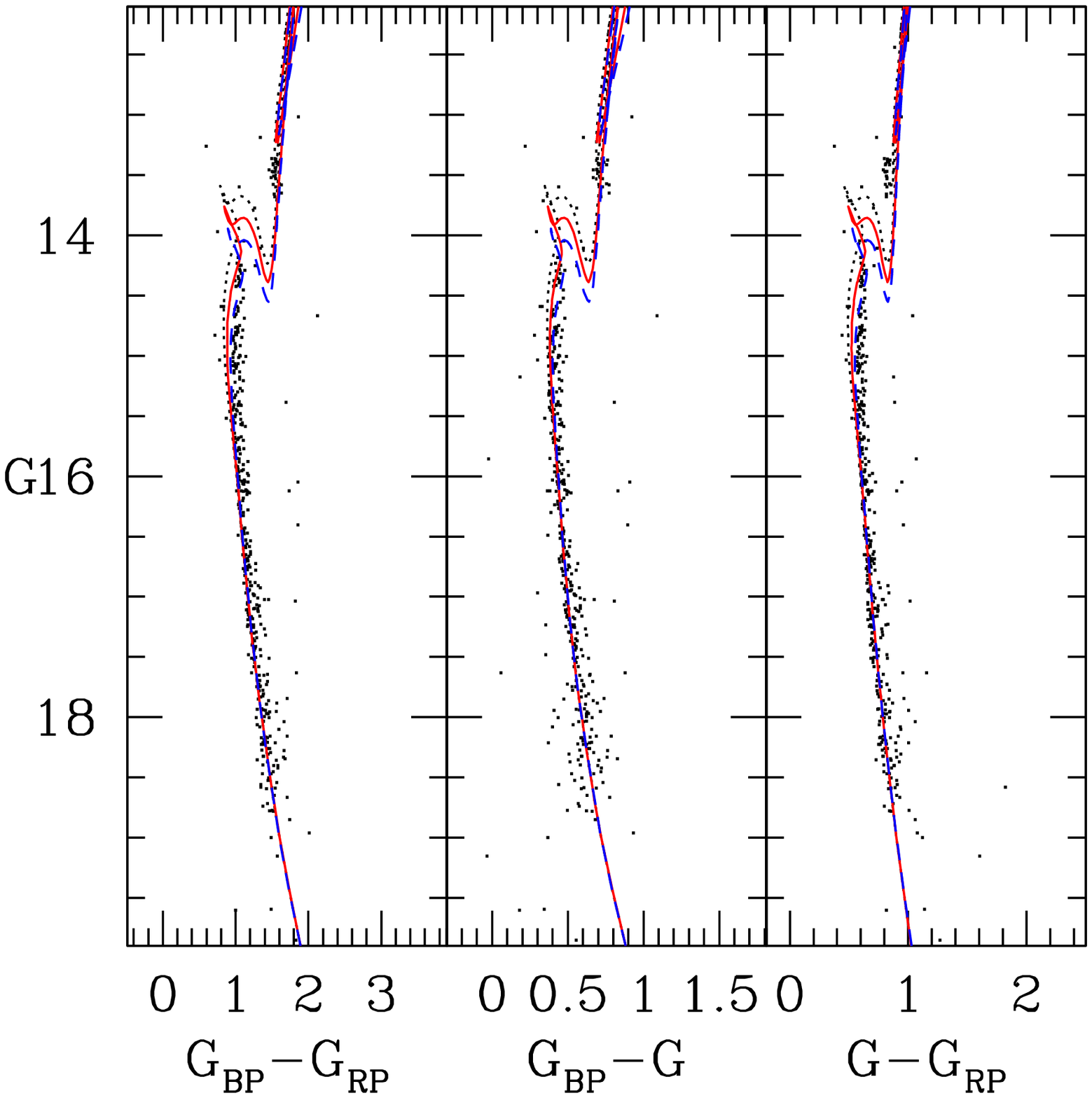}
\includegraphics[width=8.5cm, height=8.5cm]{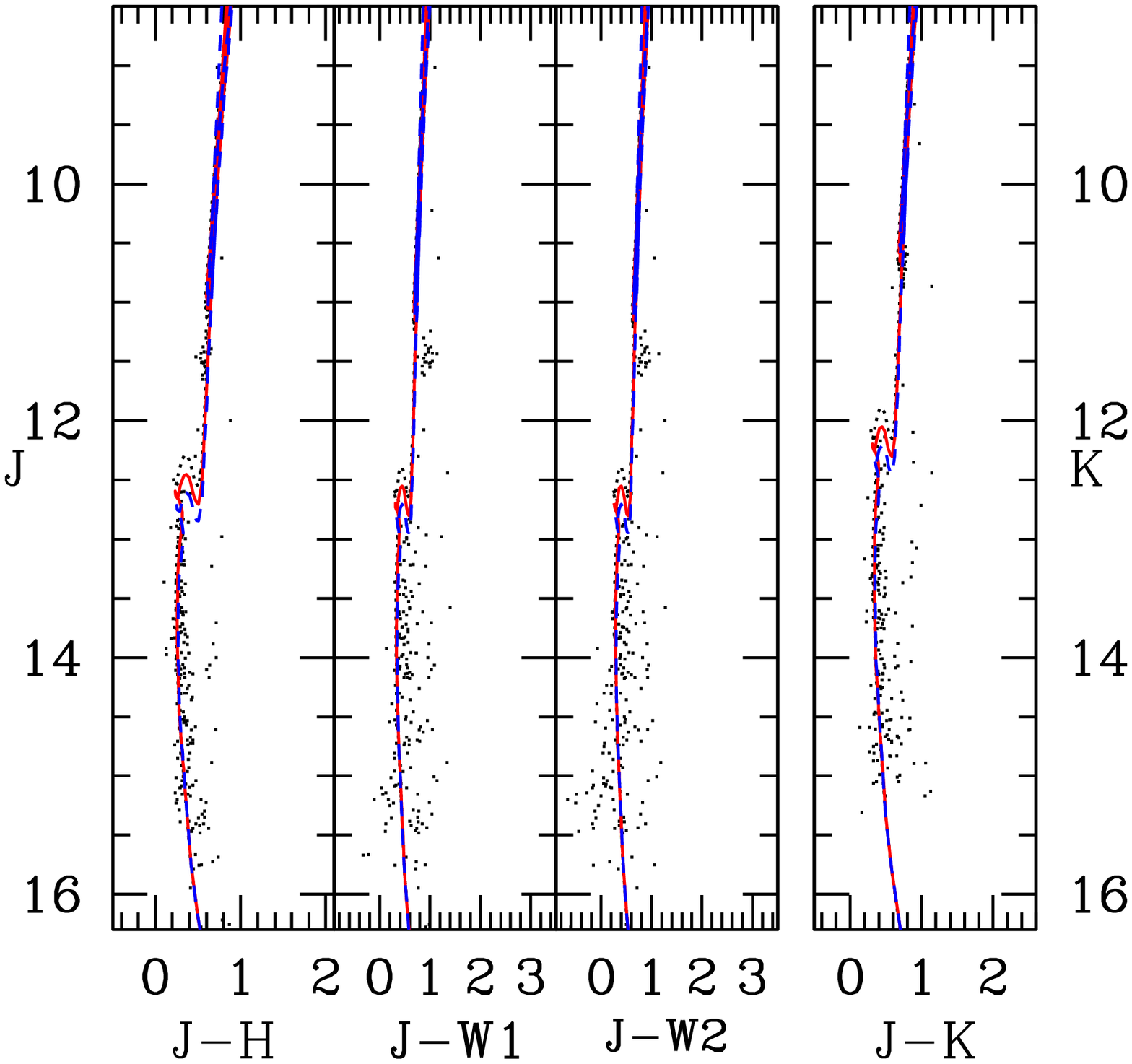}
}
\vspace{-0.5cm}\caption{The $G, (G_{BP}-G_{RP})$, $G, (G_{BP}-G)$, $G, (G-G_{RP})$, $J, (J-H)$, $J, (J-W1)$, $J, (J-W2)$ and $K, (J-K)$
color-magnitude diagrams of NGC 4337. The curves are the isochrones of log(age) $=$  9.15 ,9.20 and 9.25. These isochrones
are taken from Marigo et al. (2017).}
\label{dist1}
\end{center}
\end{figure*}

Age and distance are very important parameters to trace the structure and chemical evolution of the Galaxy using OCs
(Friel \& Janes 1993). The main astrophysical parameters (age, distance and reddening) of NGC 4337 are estimated by
fitting the solar metallicity theoretical evolutionary isochrones of Marigo et al. (2017) to the observed CMDs as shown
in Fig.~\ref{dist1}. In Fig.~\ref{dist1}, we superimposed isochrones of different log(age) values (9.15, 9.20 and 9.25)
having $Z=0.019$ in $G, (G_{BP}-G_{RP})$, $G, (G_{BP}-G)$, $G, (G-G_{RP})$, $J, (J-H)$, $J, (J-W1)$, $J, (J-W2)$ and
$K, (J-K)$ CMDs. The overall fit is good for log(age)=9.20 (middle isochrone), corresponding to $1600\pm180$ Myr. The
estimated distance modulus ($m-M_{k}$=12.40 mag) provides the heliocentric distance as $2.5\pm0.06$ kpc.

\begin{figure}
\begin{center}
\includegraphics[width=7.5cm, height=7.5cm]{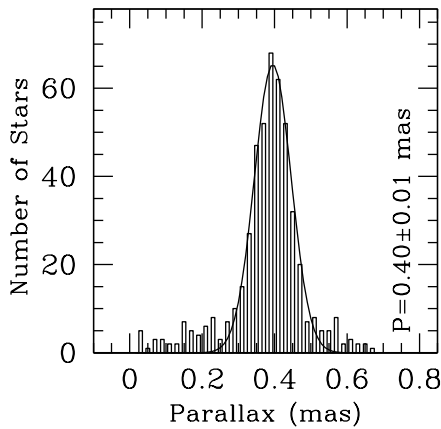}
\caption{Histogram of parallax estimation of cluster NGC 4337 using the most probable members. The Gaussian function
fit to the central bin provides mean parallax.}
\label{pllax}
\end{center}
\end{figure}

To estimate the distance of the cluster, we also used the parallax of cluster members available in Gaia~DR2 catalog.
We constructed a histogram using parallax in 0.15 mas bins as shown in Fig.~\ref{pllax} using the most probable members selected
from the cluster's VPD. By fitting the Gaussian function, mean parallax is estimated as $0.40\pm0.01$ mas which corresponds to a
distance of $2.5\pm0.07$ kpc. The calculated value of parallax is in good agreement with the value obtained by
Cantat-Gaudin et al. (2018). We obtained a similar value of distance for NGC 4337 using mean parallax and distance modulus
of the cluster.

\section{Luminosity and Mass function}

\begin{figure}
\begin{center}
\includegraphics[width=6.5cm, height=6.5cm]{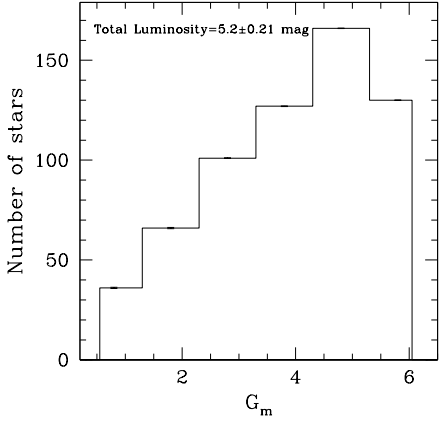}
\vspace{-0.5cm}
\caption{Luminosity function of stars in the region of NGC 4337.}
\label{lf}
\end{center}
\end{figure}

Luminosity function (LF) and Mass function (MF) are associated with each other with the well known mass-luminosity relationship.
To construct the luminosity function for the cluster NGC 4337, we used $G$ versus $(G_{BP}-G_{RP})$ CMD. Before building
the true LF, we converted the observed $G$ magnitudes of member stars into the absolute $G$ magnitudes considering the
distance modulus of the cluster. The constructed histogram of LF for NGC 4337 is shown in Fig.~\ref{lf}.

\begin{figure}
\begin{center}
\includegraphics[width=6.5cm, height=6.5cm]{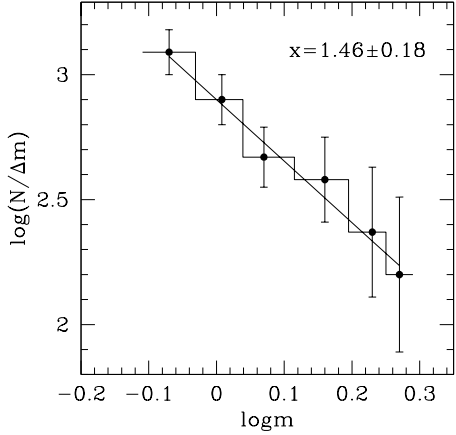}
\vspace{-0.5cm}
\caption{Mass function histogram  derived using the most probable members, where solid line indicates the power law
given by Salpeter (1955).}
\label{mass}
\end{center}
\end{figure}

MF is derived using LF, which is relative number of stars in certain interval bins of absolute magnitudes. We have used the
model given by Marigo et al. (2017) to convert the LF into MF. The resulting mass function is shown in Fig.~\ref{mass}. The mass
function slope can be derived by using the following relation\\

\begin{equation}
\log\frac{dN}{dM}=-(1+x)\log(M)+constant\\
\end{equation}

where $dN$ represents the number of stars in a mass bin $dM$ with central mass $M$ and $x$ is mass function slope.
The mass function slope for NGC 4337 is found to be $1.46\pm0.18$. The initial mass function for massive stars
($\ge$ 1 $M_{\odot}$) has been studied and well established by Salpeter (1955) and he found the value of $x$ as 1.35.

According to the Salpeter's power law, the number of stars in each mass range decreases rapidly with increasing mass.
It is noted that our investigated value of MF slope is similar to Salpeter's value. We have estimated the total mass considering
the above mass function slope within the mass range 0.75~-~2.0 $M_{\odot}$. The total cluster mass and mean mass are estimated
as $\sim$720 $M_{\odot}$ and  $\sim$1.15 $M_{\odot}$ respectively.


\section{The orbit of the cluster}

We estimated the Galactic orbit of NGC 4337 using the Galactic potential models. We adopted the method given
by Allen \& Santillan (1991) for Galactic potentials. Recently, Bajkova \& Bobylev (2016) and Bobylev et. al (2017)
refined the parameters of Galactic potential models with the help of new observational data for the galacto-centric distance
R $\sim$ 0 to 200 kpc. These potentials are given as    \\

$ \Phi_{b}(r,z) = -\frac{M_{b}}{\sqrt{r^{2} + b_{b}^{2}}} $   \\

$ \Phi_{d}(r,z) = - \frac{M_{d}}{\sqrt{r^{2} + (a_{d} + \sqrt{z^{2} + b_{d}^{2}})^{2}}}  $ \\

$ \Phi_{h}(r,z) = - \frac{M_{h}}{a_{h}} ln(\frac{\sqrt{r^{2} + a_{h}^{2}} + a_{h}}{r}) $    \\

Where  $ \Phi_{b} $ , $ \Phi_{d} $ and $ \Phi_{h} $ are the potentials of the central bulge, disc, and halo of Galaxy
respectively. $r$ and $z$ are the distances of objects from Galactic center and Galactic disc respectively. The
halo region potential is given by Wilkinson \& Evans (1999). All three potentials are axis-symmetrical, time independent
and analytical. Also, their spatial derivatives are continuous everywhere.

\subsection{Orbit calculation}

\begin{figure}
\begin{center}
\hbox{
\includegraphics[width=3.5cm, height=3.5cm]{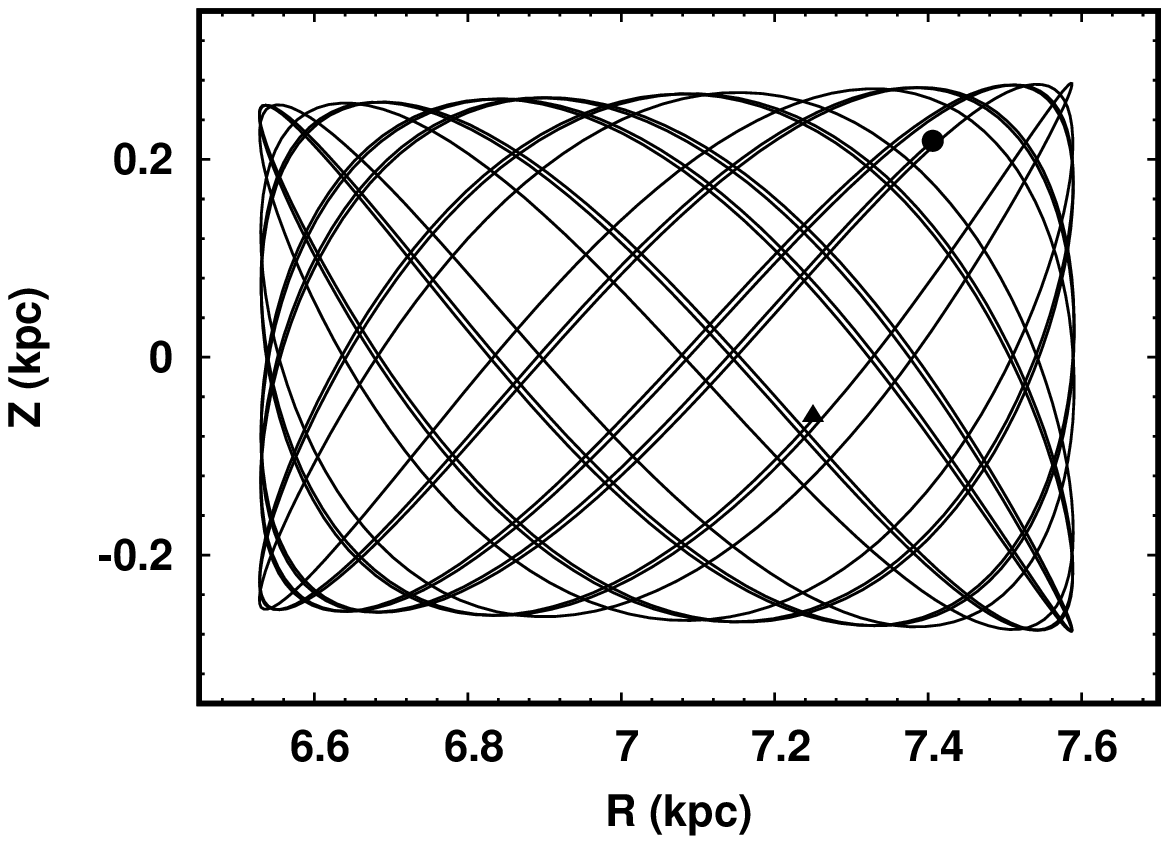}
\includegraphics[width=4.5cm, height=3.5cm]{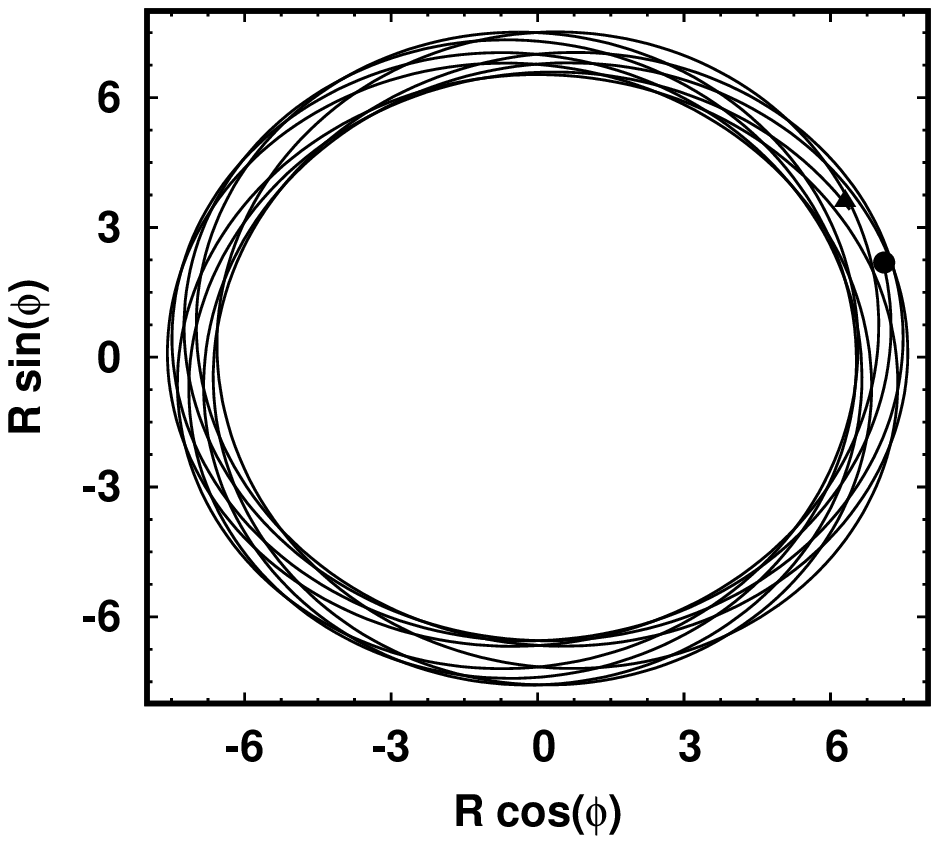}
}
\caption{Galactic orbit of the NGC 4337 estimated with the Galactic potential model described in text in the
time interval of the age of each cluster. The left panel shows a side view and the right panel shows a top
view of the cluster's orbit. The filled triangle and filled circle denotes birth and present day position of
NGC 4337 in the Galaxy. }
\label{orbit}
\end{center}
\end{figure}
To obtain orbit parameters we have used the main fundamental parameters, like central coordinates
($\alpha$ and $\delta$), mean proper motions ($\mu_{\alpha}cos\delta$, $\mu_{\delta}$), parallax angles, clusters
age and heliocentric distance ($d_{\odot}$). Mean radial velocity is estimated using 28 stars taken from the Gaia DR2 catalog.

\begin{table*}
   \centering
   \caption{Position and velocity components in the Galactocentric coordinate system. Here $R$ is the galactocentric
            distance, $Z$ is the vertical distance from the Galactic disc, $U$ $V$ $W$ are the radial tangential and the vertical
            components of velocity respectively and $\phi$ is the position angle relative to the sun's direction.
}
   \begin{tabular}{ccccccccc}
   \hline\hline
   Cluster   & $R$ &  $Z$ &  $U$  & $V$  & $W$ & $\phi$   \\
   & (kpc) & (kpc) & (km/s) &  (km/s) & (km/s) & (radian)    \\
  \hline
   NGC 4337 & 7.40 & 0.21 & $-18.85 \pm 0.47$  & $-235.40 \pm 0.44$ &  $-12.70 \pm 0.54$ & 0.29    \\
\hline
  \end{tabular}
  \label{inp}
  \end{table*}

The right-hand coordinate system is adopted to transform equatorial velocity components into Galactic-space velocity
components ($U,V,W$), where $U$, $V$ and $W$ are radial, tangential and vertical velocities respectively. In this system
the x-axis is taken positive towards Galactic-center, y-axis is along the direction of Galactic rotation and z-axis is
towards Galactic north pole. The Galactic center is taken at ($17^{h}45^{m}32^{s}.224, -28^{\circ}56^{\prime}10^{\prime\prime}$)
and North-Galactic pole is at ($12^{h}51^{m}26^{s}.282, 27^{\circ}7^{\prime}42^{\prime\prime}.01$) (Reid \& Brunthaler, 2004).
To apply a correction for Standard Solar Motion and Motion of the Local Standard of Rest (LSR), we used position coordinates of
Sun as ($8.3,0,0.02$) kpc and its space-velocity components as ($11.1, 12.24, 7.25$) km/s (Schonrich et al. 2010). Transformed
parameters in Galacto-centric coordinate system are listed in Table \ref{inp}.

Fig.\ref{orbit} shows orbit of NGC 4337. The left panel exhibits the motion in galacto-centric distance ($R$) and 
Galactic plane ($z$). The right panel shows a top view of the motion in terms of $x$ and $y$ components of Galactocentric distance.
Orbital parameters are also derived and listed in Table \ref{orpara}. Here $e$ is
eccentricity, $R_{a}$ is the apogalactic distance, $R_{p}$ is the perigalactic distance, $Z_{max}$ is the maximum distance
traveled by cluster from Galactic disc, $E$ is the average energy of orbits, $J_{z}$ is $z$ component of angular momentum
and $T$ is the time period of the cluster in the orbits.

\begin{table*}
  \centering
   \caption{Orbital parameters for NGC 4337 obtained using the Galactic potential model.
   }
   \begin{tabular}{ccccccccc}
   \hline\hline
   Cluster  & $e$  & $R_{a}$  & $R_{p}$ & $Z_{max}$ &  $E$ & $J_{z}$ & $T$   \\
           &    & (kpc) & (kpc) & (kpc) & $(100 km/s)^{2}$ & (100 kpc km/s) & (Myr) \\
   \hline\hline
   NGC 4337 &  0.004  & 7.58  & 7.51  & 0.27 & -11.92 & -17.43  & 196 \\
 \hline
  \end{tabular}
  \label{orpara}
  \end{table*}

Present analysis of the orbit shows that the cluster NGC 4337 is orbiting in a circular orbit with eccentricity $\sim$ 0.
The orbit is confined in a box of $ 7.51 < R_{gc} \leq 7.58 $ kpc. This indicates that NGC 4337 is not interacting within the inner region of the
Galaxy. The orbit is within Solar circle.

\section{Dynamical and kinematical analysis}

\begin{figure}
\begin{center}
\includegraphics[width=6.5cm, height=6.5cm]{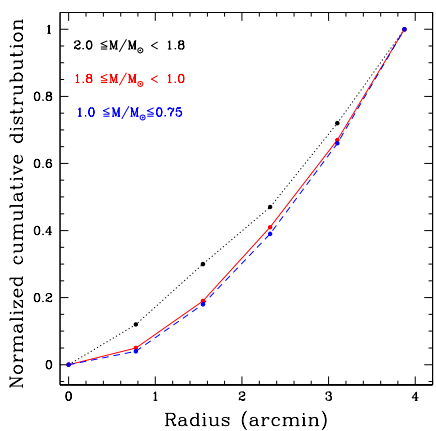}
\vspace{-0.5cm}
\caption{The cumulative radial distribution of stars in various mass ranges.}
\label{mass_seg}
\end{center}
\end{figure}

The dynamical relaxation time is the time-scale in which the cluster will lose all traces of its initial dynamic
condition (Yadav et al. 2013; Bisht et al. 2019). Because of the internal dynamics among the members,
the contraction and destruction forces make the cluster approach a Maxwellian equilibrium. During mass-segregation, massive
stars are concentrated towards the cluster core than fainter ones and this phenomenon has been reported recently for many OCs
(Piatti 2016, Zeidler et al. 2017; Dib et al. 2018, Rangwal et al. 2019, Bisht et al. 2020, Joshi et al. 2020). To see the mass-segregation
effect in the cluster NGC 4337, we have plotted the cumulative radial stellar distribution of member stars for different masses
as shown in Fig.~\ref{mass_seg}. We divided the main sequence stars in three mass ranges
i.e. 2.0$\le\frac{M}{M_{\odot}}\le$~1.8, 1.8$\le\frac{M}{M_{\odot}}\le$~1.0 and 1.0$\le\frac{M}{M_{\odot}}\le$~0.75.
Figure~\ref{mass_seg} shows the effect of mass segregation in the sense that bright stars gradually sink towards the core,
while fainter one moves away from the core. Further, we checked this signature using Kolmogrov-Smirnov $(K-S)$ test. This test
also indicates the presence of mass-segregation with a confidence level of 80$\%$ in this cluster.

%
%

During the relaxation time $t_{relax}$, which depends on both the total number of members $N$ and the cluster diameter $D$, OCs
run away to a Maxwellian stability equilibrium due to the mass-segregation phenomenon (Maciejewski \& Niedzielski 2007).
During mass-segregation in the cluster region, stars having low mass gains highest random velocities and occupy larger
volume in comparison to massive stars (Mathieu and Latham 1986).\\ 
Mathematically $t_{relax}$ can be expressed according to the
relation given by Spitzer \& Hart (1971) as;

\begin{equation}
t_{\textrm{relax}} =\frac{8.9\times 10^{5} \sqrt{N}
R_{\textrm{h}}{{  3\mathord{\left/ {\vphantom {3 2}} \right.
\kern-\nulldelimiterspace} 2} } }{\left\langle M\right\rangle
{{   1\mathord{\left/ {\vphantom {1 2}} \right.
\kern-\nulldelimiterspace} 2} } \log \left(0.4N\right)}.
\end{equation}

where $N = 624$ is the number of most probable cluster members, $R_{h}$ is the radius containing $50 \%$ of the cluster mass
(i.e. $R_{h} \simeq \frac{1}{2} \times Radius$ in pc) and $\langle M \rangle$ is the average mass of members ($\simeq 1.15 M_{\odot}$).
Thus, we estimated the value of dynamical relaxation time as $\simeq 57\pm7.55$ Myr.


Finally, we described the cluster dynamical state by computing its dynamical evolution parameter (i.e. $\tau = age / t_{relax}$).
Age of cluster is found higher than its relaxation time with $\tau\sim28$. This analysis shows that NGC 4337 is a relaxed
open cluster. All the numerical values of dynamical parameters and different times are listed in Table \ref{all}.

\subsection{Kinematical structure}

The VEPs for the most probable cluster members are computed using a computational algorithm (see Elsanhoury 2015;
Elsanhoury et al. 2018; Postnikova et al. 2020). We cross-matched cluster members with the catalog given by 
Soubiran et al. (2018). The mean radial velocity is calculated as $V_r$ (km/s) = $-$15.58 $\pm$ 0.53 km s$^{-1}$ by
using the weighted mean method. We estimated the cluster's position ($X, Y, Z$, in pc, Mihalas \& Binney 1981) and its
velocity components ($V_x$, $V_y$, $V_z$ km/s) along $x$, $y$, and $z$-axes in the coordinate system centered at 
the Sun using the formulae given by Smart (1968). Elsanhoury et al. (2016) also has been explained the equations
to estimate velocity components.

%
%
%
%
%

\subsubsection{The apex of the cluster}

\begin{figure}
\centering
\includegraphics[width=7.5cm,height=5cm]{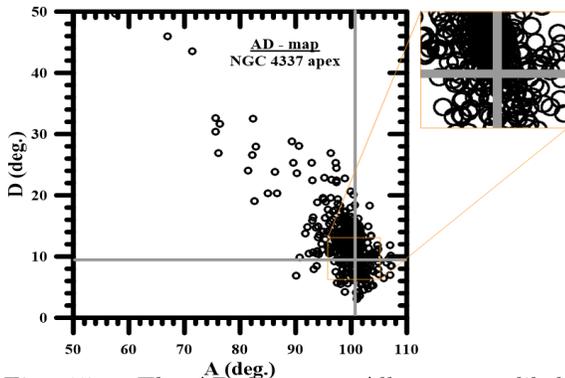} 
\vspace{-0.5cm}
\caption{The AD-diagram. All stars are likely members with membership probability higher than $50\%$.}
\label{cp2}
\end{figure}
We aim to derive apex coordinates ($A, D$) of the cluster on the celestial sphere
(Vereshchagin et al. 2014, Postnikova et al. 2020) based on 2MASS and Gaia~DR2 catalogs. The apex demonstrates
the actual direction where the object moves. To determine the cluster's apex, we adopted two different methods: (1)
the classical convergent point method  and (2) the AD-diagram method. These methods have been discussed below and the
obtained results of this analysis are listed in Table \ref{all}.\\

$\bullet$ \textit{The classical convergent point method ($A_{conv}, D_{conv}$)}

For almost a century, the classical convergent point method has been used to derive convergent point ($A_{conv}, D_{conv}$).
In later years, this method has been improved by several authors (e.g. Jones 1971; de Bruijne 1999; Galli et al. 2012). This
method assumes the same space velocities for all cluster members.

Using the above set of equations and assuming

\begin{equation}
\begin{split}
\xi = \frac{V_x}{V_z},\\
\eta = \frac{V_y}{V_z}.\\
\end{split}
\end{equation}

we get, 

\begin{equation}
a_{i} \xi + b_{i} \eta  = c_{i}.
\end{equation}

where the coefficients $a_{i}$, $b_{i}$ and $c_{i}$
for $N$ cluster members can be given as:

\begin{equation}
\begin{split}
&  a_i = \mu_\alpha^{(i)}\sin\delta_i\cos\alpha_i\cos\delta_i-\mu_\delta^{(i)}\sin\alpha_i, \\
& b_i = \mu_\alpha^{(i)}\sin\delta_i\sin\alpha_i\cos\delta_i+\mu_\delta^{(i)}\cos\alpha_i, \\
& c_i = \mu_\alpha^{(i)}\cos^2\delta_i.
\end{split}
\end{equation}

then

\begin{equation}
A_{conv}=\tan^{-1}\Big[\frac{\eta}{\xi}\Big].
\label{aconv}
\end{equation}

\begin{equation}
D_{conv}=\tan^{-1}\Big[\frac{1}{\sqrt{\eta^2+\xi^2}}\Big].
\label{dconv}
\end{equation}

The coordinates $(A_{conv}, D_{conv})$ of the cluster apex are derived from the equations \ref{aconv} and \ref{dconv}.\\

$\bullet$ \textit{The AD-diagram method ($A_{\circ}, D_{\circ}$)}\\ 

The AD-diagram is discussed in detail by Chupina et al. (2001, 2006); Vereshchagin et al. (2014) and Elsanhoury et al. (2018, 2019).
In stellar apex method, ($A, D$) of individual stars give the positions of those stars as a function of space velocity vectors
$(V_{x}, V_{y}, V_{z})$. The intersection point is called the apex ($A_{\circ}, D_{\circ}$) in equatorial coordinates,
which is given as follows:

\begin{equation}
A_{\circ}=\tan^{-1}\Big[\frac{\overline{V_y}}{\overline{V_x}}\Big].
\end{equation}

\begin{equation}
D_{\circ}=\tan^{-1}\Big[\frac{\overline{V_z}}{\sqrt{\overline{V_x}^2+\overline{V_y}^2}}\Big].
\end{equation}

Figure~\ref{cp2} shows the AD-diagram for the cluster NGC 4337. The estimated apex position
[($A_{conv}, D_{conv}$) and ($A_{\circ}, D_{\circ}$)] for the cluster NGC 4337 using above discussed methods are
presented in Table \ref{all}.

We have derived the Velocity Ellipsoid Parameters (VEPs), matrix elements ($\mu_{ij}$), direction cosines ($l_j$, $m_j$, $n_j$),
Galactic longitude of the vertex ($l_2$) and the solar elements for NGC 4337. All these parameters are listed in Table \ref{all}.
The used equations to derive these kinematical parameters are described in the appendix. In addition to this, the 3-D space
distribution of NGC 4337 cluster has been described in the appendix using likely members with membership probability higher
than 50\% as shown in Fig. \ref{ad1}.

\section{Conclusions}
\label{con}

We presented a comprehensive photometric and kinematical study of ithe poorly studied open cluster NGC 4337 using 2MASS, WISE,
APASS and Gaia~DR2 data. We have calculated the membership probabilities in the region of cluster NGC 4337 and have found 624
member stars with membership probabilities higher than $50\%$. We have used those members to derive the fundamental
parameters. We also throw some light on the dynamical and kinematical aspects of the cluster. The main points of the current
investigation are as following:

\begin{itemize}

\item The new cluster center is estimated as:
        $\alpha = 186.01\pm0.01$ deg ($12^{h} 24^{m} 2.3^{s}$)
      and $\delta = -58.12\pm0.003$ deg ($-58^{\circ} 7^{\prime} 12^{\prime\prime}$)
using the most probable cluster members. The cluster radius is estimated
      as 7.75 arcmin using a radial density profile.\

\item Based on the vector point diagram and membership probability estimation of stars, we identified 624 most probable cluster
      members for this object. The mean PMs of the cluster are estimated as $-8.83\pm0.01$ and $1.49\pm0.006$ mas yr$^{-1}$ in both
      the RA and DEC directions respectively.\

\item Distances to the cluster NGC 4337 is determined as $2.5\pm0.06$ kpc. This value is well supported by the
      distance estimated using mean parallax of the cluster.
Age is determined as $1600\pm180$ Myr by comparing the cluster CMD
with the theoretical isochrones
given by Marigo et al. (2017).
The isochrones used have a metallicity $Z=0.01$.\

\item  The mass function slope is estimated as $1.46\pm0.18$,
        which is in good agreement with the value 1.35 given by
       Salpeter (1955) for field stars in Solar neighborhood.\

\item  Mass segregation is also observed for NGC 4337.
        The K-S test indicates $80\%$ confidence level of
       mass-segregation effect.
The dynamical relaxation time is less than the cluster's age,
which demonstrates that NGC 4337 is a dynamically relaxed open cluster.\

\item The Galactic orbits and orbital parameters were estimated using Galactic potential models. We found that
NGC 4337 is orbiting in a boxy pattern.

\item The apex position $(A, D)$ is computed with convergent point
and AD-diagram methods as:
$(A_{conv}, D_{conv})$ = (96$^{\textrm{o}}$.27 $\pm$ 0$^{\textrm{o}}$.10, 13$^{\textrm{o}}$.14 $\pm$ 0$^{\textrm{o}}$.27)
and $(A_\circ, D_\circ)$ = (100$^{\textrm{o}}$.282 $\pm$ 0$^{\textrm{o}}$.100, 9$^{\textrm{o}}$.577 $\pm$ 0$^{\textrm{o}}$.323)
respectively.\

\item For ellipsoidal motion, we computed the VEPs and direction cosines ($l_{j}, m_{j}, n_{j}$) in three axes.\

\item We have evaluated the longitude of the vertex l2 with $-$0.464; non-zero value, along which the principal axis lies.\

\item Finally, we computed the projected distance $(X_{\odot}, Y_{\odot}, Z_{\odot}$ = (1.747 $\pm$ 0.004, -3.112 $\pm$ 0.005,
0.285 $\pm$ 0.002) kpc and the Solar elements $(S_\odot, l_A, b_A) = (151.985 \pm 12.328, -23^{\textrm{o}}.270, -2^{\textrm{o}}.120)$.

\end{itemize}

\begin{table*}
\small
\caption{ The kinematical parameters of NGC 4337 cluster.}
\begin{tabular}{ll}
\hline
Parameters & NGC 4337   \\ \hline
No. of members (N) & 624   \\
Distance (kpc) [photometric cal.]        &2.5 $\pm$ 0.06 \\

M$_{\textrm{total}}$ (M$_{\odot}$)&720 \\

$\langle \textrm{M} \rangle$ (M$_{\odot}$) & 1.15  \\
t$_\textrm{relax}$(Myr)   &57.00 $\pm$ 7.55  \\

$\tau$ (log t = 9.20) & 28.00 [relaxed cluster] \\

$A_{conv}$& 96$^{\textrm{o}}$.27 $\pm$ 0$^{\textrm{o}}$.10 \\
$D_{conv}$& 13$^{\textrm{o}}$.14 $\pm$ 0$^{\textrm{o}}$.27 \\

$A_\circ$& 100$^{\textrm{o}}$.282 $\pm$ 0$^{\textrm{o}}$.100 \\
$D_\circ$& 9$^{\textrm{o}}$.577  $\pm$ 0$^{\textrm{o}}$.323 \\

$\overline{V_{x}}$ (km s$^{-1}$)& -26.75 $\pm$ 0.193 \\
$\overline{V_{y}}$ (km s$^{-1}$)& 147.46 $\pm$ 12.14 \\
$\overline{V_{z}}$ (km s$^{-1}$)&25.29 $\pm$ 0.199 \\
$\overline{V_\alpha}$ (km s$^{-1}$)& -149.446 $\pm$ 12.22 \\
$\overline{V_\delta}$ (km s$^{-1}$) &22.831 $\pm$ 0.209 \\
$\overline{V_t}$ (km s$^{-1}$) &255.825 $\pm$ 15.030 \\




$\lambda_{j}$ ($\forall$ j = 1, 2, 3) (km s$^{-1}$)& 1061500, 2390.74, 237.264 \\

$\sigma_{j}$ ($\forall$ j = 1, 2, 3) (km s$^{-1}$) &1030.290, 48.895, 15.403  \\

$l_1, m_1, n_1$ & 0$^{\textrm{o}}$.873, 0$^{\textrm{o}}$.480, -0$^{\textrm{o}}$.078  \\

$l_2, m_2, n_2$ & -0$^{\textrm{o}}$.027, -0$^{\textrm{o}}$.113, -0$^{\textrm{o}}$.993  \\

$l_3, m_3, n_3$ & 0$^{\textrm{o}}$.485, -0$^{\textrm{o}}$.869,  0$^{\textrm{o}}$.085  \\

$(x_{c}, y_{c}, z_{c})$ (pc) & -1948.18, -205.177, -3150.79  \\

$V$ (km s$^{-1}$) & 151.985 $\pm$ 12.328  \\

$B_{j}$ $(\forall$ j = 1, 2, 3$)$&-4$^{\textrm{o}}$.525, -83$^{\textrm{o}}$.314, 4$^{\textrm{o}}$.910 \\

$L_{j}$ $(\forall$ j = 1, 2, 3$)$&-28$^{\textrm{o}}$.788, 103$^{\textrm{o}}$.69, -119$^{\textrm{o}}$.178 \\

$E$ (kpc)$^{\textrm{3}}$ & 3250.258  \\

$X_{\odot}$ (kpc) & 1.747 $\pm$ 0.004   \\
$Y_{\odot}$ (kpc)  & -3.112 $\pm$ 0.005  \\

$Z_{\odot}$(kpc) & 0.285 $\pm$ 0.002  \\

$S_{\odot}$ (km s$^{-1}$)&151.985 $\pm$ 12.328 \\

 $(l_{A}, b_{A})_{w.s.v.c.}$&-23$^{\textrm{o}}$.270, -2$^{\textrm{o}}$.120 \\
 $(\alpha_{A}, \delta_{A})_{w.r.v.c.}$&-79$^{\textrm{o}}$.718, -9$^{\textrm{o}}$.578 \\
 \hline
\end{tabular}
\label{all}
\end{table*}

\acknowledgements

The authors are thankful to the anonymous referee for useful comments, which improved the contents of the paper significantly.
This work has been supported by the Natural Science Foundation of China (NSFC-11590782, NSFC-11421303). Devesh P. Sariya and
Ing-Guey Jiang are supported by the grant from the Ministry of Science and Technology (MOST), Taiwan. The grant numbers are
MOST 105-2119-M-007 -029 -MY3 and MOST 106-2112-M-007 -006 -MY3. This work has made use of data from the European Space Agency
(ESA) mission GAIA processed by Gaia Data processing  and Analysis Consortium (DPAC),
(https://www.cosmos.esa.int/web/gaia/dpac/consortium).
In addition to this, It is worthy to mention that, this work has
been done by using WEBDA and the data products from the Two Micron All Sky Survey $(2MASS)$, which is a joint project of the
University of Massachusetts and the Infrared Processing and Analysis Center/California Institute of Technology, funded by the
National Aeronautics and Space Administration and the National Science Foundation (NASA).

\begin{bibliography}{}

\noindent Adams, F. C., Myers, P. C., 2001. ApJ, 553, 744.\\\\

\noindent Allen, C., Martos, M., 1988. RMxAA 16, 25.\\\\

\noindent  Balaguer-N\'{u}\~{n}ez L., Tian, K. P., Zhao, J. L., 1998, A\&AS, 133, 387. \\\\

\noindent Bisht, D., Yadav, R. K. S., Durgapal, A. K., 2017, New Astronomy, 52, 55B.\\\\

\noindent Bisht, D., Ganesh, Shashikiran., Yadav, R. K. S., Durgapal, A. K., Rangwal, Geeta. 2018, AdSpR, 61, 571B.\\\\

\noindent Bisht, D., Yadav, R. K. S., Ganesh, Shashikiran., Durgapal, A. K., Rangwal, Geeta., Fynbo, J. P. U. 2019, MNRAS, 482, 1471B.\\\\

\noindent Bisht, D., Zhu, Qingfeng., Yadav, R. K. S., Durgapal, Alok., Rangwal, Geeta., 2020, MNRAS, 494, 607-623\\\\ 

\noindent Bland-Hawthorn J., Sharma S. et al., 2019, MNRAS, 486, 1167.\\\\

\noindent Bovy, J., 2017. MNRAS, 468, L63.\\\\

\noindent Bukowiecki, L., Maciejewski, G., Konorski, P., Strobel, A. 2011, Acta Astron., 62\\\\

\noindent Cantat-Gaudin et al. 2018, A\&A, 618, A93.\\\\

\noindent Carraro, G., et al. 2014, MNRAS, 441, L36\\\\

\noindent Cardelli J. A., Clayton G. C., Mathis J. S., 1989, ApJ, 345, 245\\\\

\noindent Chumak Y. O., Platais I., McLaughlim D. E., Rastorguev A. S., Chumak, O. V., 2010, MNRAS, 402, 1841\\\\

\noindent Chupina, N. V., Reva, V. G., Vereshchagin, S. V., 2001. A\&A, 371, 115.\\\\

\noindent Chupina, N. V., Reva, V. G., Vereshchagin, S. V., 2006. A\&A, 451, 909.\\\\

\noindent de Bruijne, J. H. J., 1999. MNRAS, 306, 381.\\\\

\noindent Dias, W. S., Alessi, B. S., Moitinho, A., Lepine, J. R. D., 2002, A\&A, 389, 871.\\\\

\noindent Dib S., Schmeja S., Parker R. J., 2018, MNRAS, 473, 849\\\\

\noindent Elsanhoury, W. H., Sharaf, M. A., Nouh, M. I., Saad, A. S.,  2013. The open astronomy journal, 6, 1.\\\\

\noindent Elsanhoury, W. H., 2015. Astrophysics, 58, 522.\\\\

\noindent Elsanhoury, W. H. et al. 2016, NA, 49, 32E\\\\

\noindent Elsanhoury, W. H., 2016b. Astrophysics, 59, 246.\\\\

\noindent Elsanhoury, W. H. et al. 2018. Astrophysics and space science, 363, 58.\\\\

\noindent Elsanhoury, W. H. and Nouh, M. I., 2019. NA, 72, 19.\\\\

\noindent Friel E. D., Janes K. A., 1993, A\&A, 267, 75\\\\

\noindent Gaia Collaboration et al., 2016a, A\&A, 595, A1.\\\\

\noindent Gaia Collaboration et al., 2016b, A\&A, 595, A2.\\\\

\noindent Gaia Collaboration et al., 2018a, A\&A, 616, A1.\\\\

\noindent Gaia Collaboration et al., 2018b, A\&A, 616, A11.\\\\

\noindent Galli, P. A. B., Teixeira, R., Ducourant, C., Bertout, C., Benevides-Soares, P., 2012, A\&A. 538, A23\\\\

\noindent Girard, T. M., Grundy, W. M., Lopez, C. E., \& van Altena, W. F. 1989, AJ, 98, 227\\\\

\noindent Heden A., Munari U., 2014, Contrib. Astron. Obs. Skalnate Pleso, 43, 518\\\\

\noindent Heden A. A., Levine S., Terrell D., Welch D. L., 2016, AAS, 225, 16H\\\\

\noindent Jones, B. F., AJ, 1971, 76, 470\\\\

\noindent Jeffries, R.D., Thurston, M.R., Hambly, N.C. 2001, A\&A, 375, 863.\\\\

\noindent Joshi Y. C., Maurya J., John A. A., Panchal A., Joshi, S., Kumar, B., 2020, MNRAS, 492, 3602\\\\

\noindent Kharchenko N. V., Piskunov A. E., Schilbach E., Röser S., Scholz R. D., 2013, A\&A, 558, A53\\\\

\noindent King I., 1962, AJ, 67, 471.\\\\

\noindent Lada, C. J., Lada, E. A., 2003, ARAA, 41, 57.\\\\

\noindent Liu, J. C., Zhu, Z., Hu, B., 2011, A\&A, 536, A102.\\\\

\noindent Maciejewski, G., Niedzielski, A., 2007, A\&A, 467, 1065\\\\

\noindent Marigo, P. et al. 2017, ApJ, 835, 77\\\\

\noindent Mathieu, R. D., Latham, D. W., 1986, AJ, 92, 1364\\\\

\noindent Mihalas, D., Binney, J., 1981, Galactic astronomy:structure and kinematics/2nd edition.\\\\

\noindent Morgan, D. H., Nandy, K. 1982, MNRAS, 199, 979.\\\\

\noindent Ogorodnikov KF. Dynamics of stellar systems. Oxford:Pergamon 1965.\\\\

\noindent Perottoni, H. D. et al., 2019, MNRAS, 486, 843.\\\\

\noindent Piatti A. E., 2016, MNRAS, 463, 3476\\\\

\noindent Postnikova E. S., Elsanhoury W. H. et al., 2020, RAA, Vol. 20, No. 2, 16.\\\\

\noindent Peterson, C. J. \& King, I. R., 1975, AJ, 80, 427.\\\\

\noindent Rangwal, G., Yadav, R. K. S., Durgapal, A., Bisht, D., Nardielo D., 2019, MNRAS, 490, 1383\\\\

\noindent Reid M. J., Brunthaler A.2004, ApJ, 616, 872\\\\

\noindent Roser et al. 2011, A\&A, 531, A92.\\\\

\noindent Roser, S., Schilbach, E., 2019, A\&A, 627, A4.\\\\

\noindent Smart, W. M., Stellar kinematics. 1968.\\\\

\noindent Soubiran et al. 2018, A\&A, 619, A155.\\\\

\noindent Seleznev, A. F. et al., 2017, MNRAS, 467, 2517.\\\\

\noindent Salpeter, E .E., 1955. ApJ 121, 161.\\\\

\noindent Sanders, W. L., 1971, A\&A, 14, 226 \\\\

\noindent Sariya, D. P., Jiang, I.-G., \& Yadav, R. K. S. 2017, AJ, 153, 134.\\\\

\noindent Sariya, D. P., Jiang, I.-G., \& Yadav, R. K. S. 2018a, Astrometry and Astrophysics in the Gaia Sky, 330, 251. \\\\

\noindent Sariya, D. P., Jiang, I.-G., \& Yadav, R. K. S. 2018b, RAA, 18, 126.\\\\

\noindent Spitzer, L., Hart, M., 1971. ApJ 164, 399.\\\\

\noindent Skrutskie, M. F., Cutri, R. M., Stiening, R., et al. 2006, AJ, 131, 1163.\\\\

\noindent Sanders W. L. 1971, A\&A, 14. 226. \\\\

\noindent Schonrich, Ralph., Binney, James., Dehnen, Walter. 2010, MNRAS, 403, 1829S\\\\

\noindent Vasilevskis S., Klemola, A., Preston, G., 1958, AJ, 63, 387.\\\\

\noindent Vereshchagin, S. V., Chupina, N. V., Sariya, D. P., Yadav, R. K. S., \& Kumar, B. 2014, New Astron., 31, 43.\\\\

\noindent von Hoerner S., 1957, ApJ, 125, 451\\\\

\noindent Wang, H. F. et al., 2019, arxiv e-prints, arxiv:1905.11944\\\\

\noindent Wright E. L., Eisenhardt P. R. M., Mainzer A. K., Ressler M. E., Cutri R. M., Jarrett T., Kirkpatrick J. D.,
          Padgtt D., 2010, AJ, 140, 1868\\\\

\noindent Yadav, R. K. S., Sariya, D. P., \& Sagar, R., 2013, MNRAS, 430, 3350.\\\\

\noindent Zeidler P., Nota A., Grebel E. K., Sabbi E., Pasquali A., Tosi M., Christian C., 2017, AJ, 153, 122

\end{bibliography}

\appendix

\section{The Velocity Ellipsoid Parameters (VEPs)}

\begin{figure*}
\centering
\includegraphics[width=9.5cm,height=9.5cm]{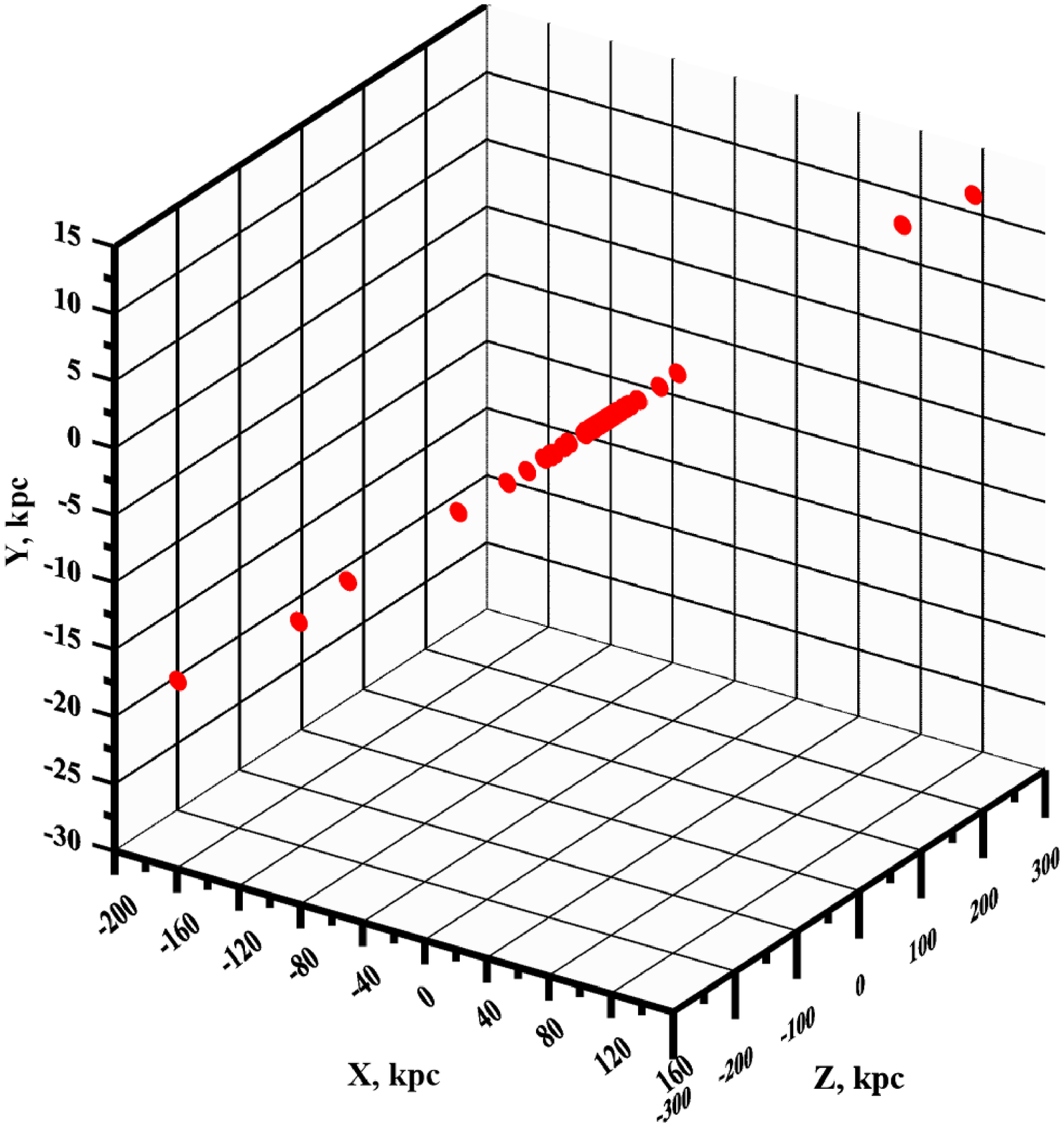}
\vspace{-0.5cm}
\caption{The 3-D space distribution of NGC 4337 cluster using likely members with membership probability higher than $50\%$.}
\label{ad1}
\end{figure*}

To compute VEPs for NGC 4337, we follow a computational algorithm (Elsanhoury et al. 2013, 2018; Elsanhoury 2015;
Postnikova et al. 2020). Depending on the matrix that controls the eigenvalue problem for the velocity
ellipsoid, we established an analytical expressions of some parameters in terms of the matrix elements $mu_{ij}$.

$\bullet$ \textit{The $\sigma_j$ parameters, $\forall j=1, 2, 3$}\\

The velocity dispersion ($\sigma_1, \sigma_2, \sigma_3$) are given here as:

\begin{equation}
\sigma_j=\sqrt{\lambda_j}.
\end{equation}\\

$\bullet$ \textit{The $l_j, m_j,$ and $n_j$ parameters, $\forall j=1, 2, 3$}\\

The $l_j, m_j,$ and $n_j$ are the direction cosines for the above eigenvalue problem. Thus, we have:\\

\begin{equation}
\begin{split}
& l_j =\frac{\mu_{22} \mu_{33} -\sigma_j^2\left(\mu_{22} +\mu_{33} -\sigma_j^2\right)-\mu_{23}^2}{D_j},\; j=1,2,3,\\
& m_j =\frac{\mu_{23} \mu_{13} -\mu_{12} \mu_{33} +\sigma_{j}^{2} \mu_{12}}{D_j}, \; j=1,2,3, \\
& n_j =\frac{\mu_{12} \mu_{23} -\mu_{13} \mu_{22} +\sigma_{j}^{2} \mu_{13}}{D_j}, \; j=1,2,3.
\end{split}
\end{equation}

where\\
$l_{j}^2+m_{j}^2+n_{j}^2 = 1$\\

and
\begin{center}
$D_j^2=\left(\mu_{22} \mu_{33} -\mu_{23}^{2} \right)^{2}+\left(\mu_{23} \mu_{13} -\mu_{12} \mu_{33} \right)^{2}$\\
$\!+\!\left(\mu_{12} \mu_{23} \!-\!\mu_{13} \mu_{22} \right)^{2}\!+\!2[\left(\mu_{22}\! +\!\mu_{33} \right)\left(\mu_{23}^{2}\! -\!\mu_{22} \mu_{33} \right)$\\
$\!+\!\mu_{12} \left(\mu_{23} \mu_{13} \!-\!\mu_{12} \mu_{33} \right)\!+\!
\mu_{13} \left(\mu_{12} \mu_{23}\! -\!\mu_{13} \mu_{22} \right)]\sigma_j^2$\\
$\!+\!\left(\mu_{33}^{2} +4\mu_{22} \mu_{33} +\mu_{22}^{2} \!-\!2\mu_{23}^{2} +\mu_{12}^{2} +\mu_{13}^{2} \right)\sigma_j^4$\\
$-2\left(\mu_{22} +\mu_{33} \right)\sigma_j^6 +\sigma_j^8$.
\end{center}

$\bullet$ \textit{The center of the cluster ($x_c, y_c, z_c$)}\\

The center of the cluster is derived by the simple method of determining the equatorial coordinates
of the center of mass of $N_i$ number of discrete objects,

\begin{center}
$x_c=\frac{\sum\limits_{i=1}^{N}\left(r_i\cos\alpha_i\cos\delta_i\right)}{N}$, \\
$y_c=\frac{\sum\limits_{i=1}^{N}\left(r_i\sin\alpha_i\cos\delta_i\right)}{N}$, \\
$z_c=\frac{\sum\limits_{i=1}^{N}\left(r_i\sin\delta_i\right)}{N}.$
\end{center}

$\bullet$ \textit{The velocity V of the cluster}\\

As a function of radial velocity $V_{r}$, the velocity of the cluster can be written in the following form:
\begin{equation}
V = \Big[\frac{\sum_{i=1}^{N}V_{r}^{(i)} \cos\lambda_{i}}{\sum_{i=1}^{N} \cos^2\lambda_{i}}\Big].
\end{equation}
where, $\lambda_{i}$ is the angular distance of the star from the vertex and

\begin{equation}
\lambda_{i}=cos^{-1}[sin\delta_{i}sinD+cos\delta_{i}cosD cos(A-\alpha_{i})]
\end{equation}

$\bullet$ \textit{The $L_{j}$ and $B_{j}$ parameters}\\

The $L_{j}$ and $B_{j}$ $(\forall j = 1, 2, 3)$ are Galactic longitude
and the Galactic latitude of the directions which correspond
to the extreme values of the dispersion. Then,

\begin{equation}
L_{j} =tan^{-1}\Big[\frac{-m_{j}}{l_{j}}\Big],
\end{equation}
\begin{equation}
B_{j} =sin^{-1}\Big[n_{j}\Big].
\end{equation}

$\bullet$ \textit{The E parameter}\\
The $E$ parameter represents the volume of the ellipsoid,\\
i.e.
\begin{equation}
E =\frac{4}{3} \pi \sigma_{1} \sigma_{2} \sigma_{3}.
\end{equation}

$\bullet$ \textit{The Solar elements}\\

Considering a group with spatial velocities ($\overline{U}, \overline{V}, \overline{W}$), the components of the Sun's velocities
are $U_{\odot}=\overline{-U}$, $V_{\odot}=\overline{-V}$ and $W_{\odot}=\overline{-W}$. Therefore, the solar elements with
spatial velocities considered (w.s.v.c.) have been estimated using the equations given by Elsanhoury et al. (2016).

By considering the position along x, y, and z axes in the coordinate system whose centered at the Sun. Then the
Sun's velocities with respect to this same group and referred to the same axes are given as; $X_{\odot}^{\circ}=-\overline{V}_{x}$,
$Y_{\odot}^{\circ}=-\overline{V}_{y}$ and $Z_{\odot}^{\circ}=-\overline{V}_{z}$. Then after, we have obtained the Solar elements
with radial velocities considered (w.r.v.c.) as;

\begin{equation}
S_{\odot}=\sqrt{({X_{\odot}}^{\circ})^{2}+({Y_{\odot}}^{\circ})^{2}+({Z_{\odot}}^{\circ})^{2}}
\end{equation}

\begin{equation}
\alpha_{A}=tan^{-1}(\frac{Y_{\odot}^{\circ}}{X_{\odot}^{\circ}}) 
\end{equation}

\begin{equation}
\delta_{A}=tan^{-1}(\frac{Z_{\odot}^{\circ}}{\sqrt{(X_{\odot}^{\circ})^{2}+(Y_{\odot}^{\circ})^{2}}})
\end{equation}

where $l_{A}$ and $\alpha_{A}$ are the Galactic longitude and right ascension while $b_{A}$ and
$\delta_{A}$ are the Galactic latitude and declination of the Solar apex. $S_{\odot}$ is considered
as the absolute value of the Sun's velocity to out groups under investigations.

$\bullet$ \textit{NGC 4337 in 3-D}\\

In the broader terms, our Galaxy comprises two main structural elements: a spheroidal component and a disk.
Each of these contains different characteristic of stellar and non-stellar populations, and they have
different compositions, kinematics, and dynamical properties and evolutionary histories. The spheroidal component 
can be considered as being approximately axially symmetric system in which the stars are distributed at random orbits.
The disk is an extremely thin (about 200 pc thick), flat system extended in the Galactic plane from the Galactic center
to a radius of about 25 to 30 kpc (Mihalas \& Binney 1981). We expect to find one axis of the velocity ellipsoid of stars
in the Galactic plane pointing exactly towards the Galactic center. This expectation is compatible here by analysis of VEPs
(i.e. l2 with our series of articles (e.g. Elsanhoury et al., 2015, 2016). Thus, we conclude that is that
the longitude of the vertex l2 differs slightly from zero.

Figure \ref{ad1} presents 3-D plot of our cluster.
Here, we can infer that the deviation into space may be along the Galactic center, which may lead the system to behave
in space like a cascaded system. This causes a gradual transformation of the cluster into a stream deviating into ordered ring
structures stretched around the Galactic center (Perottoni et al. 2019; Wang et al. 2019).

\end{document}